\documentclass[%
 reprint,
 amsmath,amssymb,
 aps,
 prd,
]{revtex4-2}

\usepackage{graphicx}
\usepackage{dcolumn}
\usepackage{bm}
\usepackage{lipsum} 
\usepackage{url} 
\usepackage{orcidlink}
\usepackage{diagbox} 
\usepackage{booktabs} 

\begin{document}

\preprint{APS/123-QED}

\title{Accounting for Tidal Deformability in Binary Neutron Star Template Banks}

\author{Lorenzo Piccari \orcidlink{0009-0000-0247-4339}}

\author{Francesco Pannarale \orcidlink{0000-0002-7537-3210}}

\affiliation{Dipartimento di Fisica, Università di Roma ``Sapienza'', Piazzale A. Moro 5, I-00185, Rome, Italy\\
INFN Sezione di Roma, Piazzale A. Moro 5, I-00185, Rome, Italy}

\date{\today}

\begin{abstract}
Modelled searches for gravitational waves emitted by compact binary coalescences currently filter the data with template signals that ignore all effects related to the physics of dense-matter in neutron stars interiors, even when the masses in the template are compatible with a binary neutron star or a neutron star--black hole binary source. The leading neutron star finite-size effect is an additional phase contribution due to tidal deformations induced by the gravitational coupling between the two inspiralling objects in the binary.  We show how neglecting this effect in the templates reduces the search sensitivity close to the detection threshold.  This is particularly true for binary neutron stars systems, where tidal effects are larger.  In this work we therefore propose a new technique for the construction of binary neutron star template banks that accounts for neutron star tidal deformabilities as degrees of freedom of the parameter space to be searched over.  A first attempt in this direction was carried out by Harry \& Lundgren [Physical Review D 104, 043008 (2021)], who proposed to extract randomly the tidal deformabilities of the stars over a uniform interval, regardless of the binary neutron star component masses.  We show that this approach yields 33\% additional templates with respect to the equivalent point-like template bank.  Our proposed approach, instead, adopts a more physically motivated tidal deformability prior with a support that is informed by the value of the neutron star mass and compatible with the neutron star equation of state constraint provided by the observation of GW170817.  This method significantly reduces the needed additional templates to 8.2\%.
\end{abstract}

\maketitle

\section{\label{sec:level1}Introduction}

The first direct detection of a gravitational-wave (GW) signal was achieved by the network of two LIGO detectors on September 14, 2015.  The event, labelled GW150914 \cite{GW150914}, was associated to a compact binary coalescence (CBC) and more specifically to a binary black hole (BBH) merger; it ushered the era of GW astronomy, which constitutes a new way to probe our Universe.
Since then, during the first three Observing Runs (O1, O2, and O3) and the first part of the ongoing Fourth Observing Run (O4), the LIGO-Virgo-KAGRA (LVK) Collaboration uncovered \( 218\) GW signals \cite{GWTC_4_intro, GWTC_4_methods, GWTC_4_results, GWTC_4_GWOSC, Aasi_2015, Acernese_2015, KAGRA:2020tym}. Additionally, at the time of writing, the LVK announced the observation of the GW250114 event \cite{GW250114_discovery}, in addition to releasing more than 100 low-latency public alerts during the rest of the ongoing run \footnote{\url{https://gracedb.ligo.org}.}.
All observations registered so far were most likely emitted by BBHs, with only six compatible with the presence of at least one neutron star (NS) in the coalescing binary \cite{GW170817, Abbott_2020, Abbott_2021, Abac_2024}.
Among these, the most relevant event is undoubtedly GW170817, the first detected binary neutron stars (BNS) signal and so far the only one observed along with its electromagnetic counterpart \cite{GW170817, GW170817_prop, GW170817_multi, GW170817_grb, Gw170817_cosmo, Gw170817_GR}.
 
One of the key scientific goals of detecting GW signals from coalescences involving NSs is to probe the nuclear equation of state (EoS) in the density range \( \rho \sim 2 \div 8 \cdot 10^{14}\)\,g/cm\(^3\) typical of matter in NSs interiors, which is currently underconstrained \cite{Benhar:2024qcw, chatziioannou_2024}. The NS internal structure has a direct imprint on the GW signature of BNS and neutron star--black hole (NSBH) systems \cite{Chatziioannou_2020}. 
The sensitivity of current, second-generation detectors is not sufficient to observe the post-merger of these sources \cite{GW170817_postmerg}.
Consequently, the main observable finite-size effect is the deviation they introduce in the inspiral phase of the gravitational waveform \cite{Flanagan_2008}. The dominant contribution is given by the mutual tidal gravitational interaction between the two compact binary constituents. Such interaction induces a tidal deformation (a bulge) on the NSs in the system, which gravitationally affects the orbital motion of the binary and consequently the phase evolution of the emitted GW waveform.
At leading order, the parameter that quantifies how easily a NS is deformed by an external tidal field is the \textit{tidal deformability} \( \lambda \), defined as the ratio between the quadrupole moment $Q_{ij}$ induced in the mass distribution of the star and the external tidal field $\mathcal{E}_{ij}$ inducing it: \(\lambda = -Q_{ij} / \mathcal{E}_{ij}\). The value of $\lambda$ is proportional to the quadrupolar, electric-type tidal Love number $k_2$, hence it depends on the EoS model assumed for the NS matter, e.g., \cite{Damour:2009vw, Chatziioannou_2020}. Inferring the value of $\lambda$ for the two NSs in GW170817 has already enabled ruling out some of the stiffest EoS models, that is, those supporting less compressible NS matter and therefore less compact NSs at a fixed NS mass value \cite{GW170817_prop}.

Tidal deformability is currently not accounted for in the data analysis technique primarily used to detect CBC signals, namely \textit{matched filtering} \cite{LVKguide, pycbc_guide, Owen_1996}. This is the optimal technique to extract signals buried in much larger noise, if the noise is dominated by a Gaussian stationary component. Matched filtering consists in convolving the detector strain output data with a theoretically predicted waveform, referred to as \textit{template}, that represents the expected signal.
Since the GW signal potentially present in the detector data depends on several parameters that are unknown at the time of observation, the data are cross-correlated with a discrete set of templates chosen to cover the target parameter space of physical CBC sources and identify the template which correlates best with the data.
The ensemble of templates used for a search is called \textit{template bank}.
The efficiency of matched-filtering searches is inextricably linked to the precision of the waveform model used to generate the templates and to the efficiency of the algorithm employed for distributing templates across the parameter space.

Current CBC searches based on matched-filtering treat compact objects in the binary as point particles, neglecting any kind of effect due to the internal structure of NSs. Equivalently, template banks are built assuming a vanishing NS tidal deformation imprint on the GW signal emitted by inspiralling BNS and NSBH systems.
This choice is made in light of the fact that such effects remain small until the final stages of the inspiral, just before the merger. For BNSs, in particular, this occurs at frequencies close to the upper limit of the sensitivity band of second-generation detectors, making these effects challenging to observe \cite{Chatziioannou_2020}.
However, at lower frequencies, farther from the merger, tidal deformation effects, although small, are not completely negligible and ignoring them results in a reduction of search sensitivity with respect to signals close to the detection threshold \cite{Hinderer_2009, Cullen_2017, Harry_2018}.

Resolving this shortcoming requires the construction of \textit{tidal-template banks}, i.e., template banks accounting for NS tidal deformation effects.  The challenge is to build such template banks while also keep their size under control, in order to limit the computational cost of hypothetical searches using them. Indeed, computational cost is one of the main bottlenecks of template bank-based searches. This is directly linked to the number of templates in the bank, which is proportional to the size and dimensionality of the parameter space to be covered \cite{Allen_2021}.
Usually, this is limited to the masses and the spin components aligned to the orbital angular momentum of the binary. 
Dropping the assumption of negligible tidal deformation effects thereby increases the dimensionality of the target parameter space by one or two for BNS and NSBH systems, respectively.
Reducing as much as possible the number of additional templates required to cover this augmented target space is a key aspect when building a tidal-template bank.

This work proposes a new technique to account for tidal deformabilities in the construction of tidal-template banks in an efficient way. This is based on including the constraints on the NS EoS provided by the observation of GW170817 and its EM counterpart \cite{GW170817_prop, Radice_2018, Radice_2019} in a stochastic template placement algorithm \cite{Harry_2009, kacanja_2024}. 
We compare our method to the work by \citet{Harry_Lundgren}, who first proposed accounting for tidal deformabilities in building template banks.  We demonstrate that our proposal substantially reduces the number of additional templates.  At the same time, better performances in terms of signal uncovering can be obtained against BNS and NSBH signals, compared to point-like template banks.
Our technique was implemented adopting tools of the software package \texttt{PyCBC} \cite{PyCBC}, which we used for all the results presented is this work.

The remainder of this paper is structured as follows. In Section \ref{sec:II}, after a brief review of the process of validating a template bank, we present the results of tests carried out on a standard, i.e., point-particle, template bank demonstrating that neglecting NS tidal deformability effects does indeed cause a reduction in search sensitivity. In Section \ref{sec:III} we present our new technique and discuss the results obtained when using it to construct a BNS tidal-template bank.

\section{The effect of neglecting Neutron stars tidal deformations}\label{sec:II}

In this section we discuss the results obtained by testing a template bank built treating the binary constituents as point-particles and adopting a publicly available noise-power spectral density that is representative of the LIGO O4 design sensitivity ({\ttfamily aligo\_O4low.txt} \footnote{\url{https://dcc.ligo.org/LIGO-T2200043/public}}); this specific noise-power spectral density is adopted throughout the paper.
The ranges of physical parameters covered by the bank are the following. For component masses: $ 1 M_\odot \leq m_1 \leq 499 M_\odot $ and $ 1 M_\odot \leq m_2 \leq 250 M_\odot $, with additional constraints on the mass ratio, $1 \leq q=m_1/m_2 \leq 97.989$, and the total mass, $ m_1+m_2 \leq 500 M_\odot$. Assuming non-precessing binaries, the dimensionless spin magnitudes vary in the ranges $-0.997 \leq \chi_{BH} \leq 0.997$ and $-0.05 \leq \chi_{NS} \leq 0.05$, for black holes (BHs) and NSs, respectively. The boundary mass to distinguish between a BH and a NS is set to $2.0 M_\odot$. The template bank was generated using a hybrid geometric-random method  \cite{Roy:2017qgg, Roy:2017oul}, the waveform approximant \texttt{SEOBNRv5\_ROM} \cite{Buonanno_1999, Buonanno_2007, Bohe_2017, Pompili_2023} and a minimum match of $0.965$.

We obtain clear evidence of the negative impact that neglecting tidal deformations has on the performance of the template banks. The template bank characterization procedure and the results are reported in Sec.\,\ref{sec:IIA} and Sec.\,\ref{sec:IIB}, respectively.

\subsection{Template bank effectualness}\label{sec:IIA}

The main goal in the construction of a template bank is to cover the intrinsic parameter space of the target sources with the least number of templates, while allowing for a predetermined fraction of lost signals.  This parameter space can be visualized as a continuous manifold that needs to be efficiently sampled to maximize the signal-to-noise ratio (SNR) of any possible incoming signal. In this framework we define the \textit{match} \( \mathcal{M} \) between two waveforms \(h_i \equiv h_i(t)\) and \(h_j \equiv h_j(t)\) as:
\begin{equation}\label{eqn:match}
  \mathcal{M} \left ( h_i, h_j \right ) = \max_{\{\theta_{\rm ext}\}}{ \frac{\left ( h_i | h_j \right )}{\sqrt{\left ( h_i | h_i \right ) \left ( h_j | h_j \right )}}}\,,
\end{equation}
where \(\left ( h_i | h_j \right )\) is the noise-weighted scalar product defined as:
\begin{equation}\label{eqn:overlap}
\left(h_i \middle| h_j\right) = 4\mathcal{R} \int_{f_{\rm low}}^{f_{\rm high}} \frac{\tilde{h}_i^*(f) \tilde{h}_j(f)}{S_n(f)} \, df \,,
\end{equation}
with the integration interval \( \left [ f_{\rm low}, f_{\rm high} \right ] \) determined by the bandwidth of the detector data and \(S_n\) denoting the one-sided noise-power spectral density of the detector. The maximization in Eq.\,\eqref{eqn:match} is performed analytically over the extrinsic parameters \( \{\theta_{\rm ext}\} \) of the signal, namely, the source luminosity distance, sky-location, and inclination angle, as well as the polarization angle, and the time and phase of coalescence. This analytical maximization is made possible by the fact that  these parameters enter the waveform as a constant shift in phase and time, and a constant scaling in the amplitude. 
As a consequence of the normalizations in the denominator of Eq.\,\eqref{eqn:match}, the value of \( \mathcal{M}\) spans the range $[0,1]$.

In template placement algorithms, the match between templates is calculated to spread them efficiently over the parameter space. Template banks used in searches are designed with a target minimum match value between templates and any possible waveform from the target parameter space. The value of the minimum match is specifically chosen to find a viable balance between the number of required templates for coverage and the bank's performance in discovering signals. Assuming that every signal is recovered by at least one template in the bank with a match of $\bar{\mathcal{M}}$, then the requirement on the loss in detection volume is $1 - \bar{\mathcal{M}}^3 = 0.10$ when accepting a signal loss of 10\% \emph{at most}. In this case, the minimum match is $0.965$.

Once a template bank is built, its \textit{effectualness} must be verified before it can be used in real searches. This means ensuring that the bank is able to detect any potentially incoming signal within the limit of maximum SNR loss set during the template placement. We will refer to this validation process as \textit{bank verifier}, or \textit{effectualness test}. 
A bank verifier consists in a Monte Carlo simulation. A set of simulated signals is generated randomly, extracting the source physical parameters from the target parameter space covered by the bank.  Then, the \textit{Fitting Factor} (FF) of the template bank for each simulated signal, or \textit{injection}, is calculated.  This is defined as the match between the injection, \(i(t)\), and the templates maximized over the entire template bank, that is,
\begin{equation}\label{eqn:FF}
 {\rm FF}(i) = \max_{h_j \in \{{\rm bank}\}} \mathcal{M}(i, h_j)\,.
\end{equation}
This procedure maps the FF of the bank over the target parameter space in order to exclude the presence of low-performing regions. The bank is considered to be validated if less than  1\% of the total number of injections in a set is recovered with a FF lower than the minimum match threshold used for the bank construction.

\subsection{Validation against tidal deformations effects}\label{sec:IIB}

\begin{figure}[t!]
    \centering
    \includegraphics[width=1.0\linewidth]{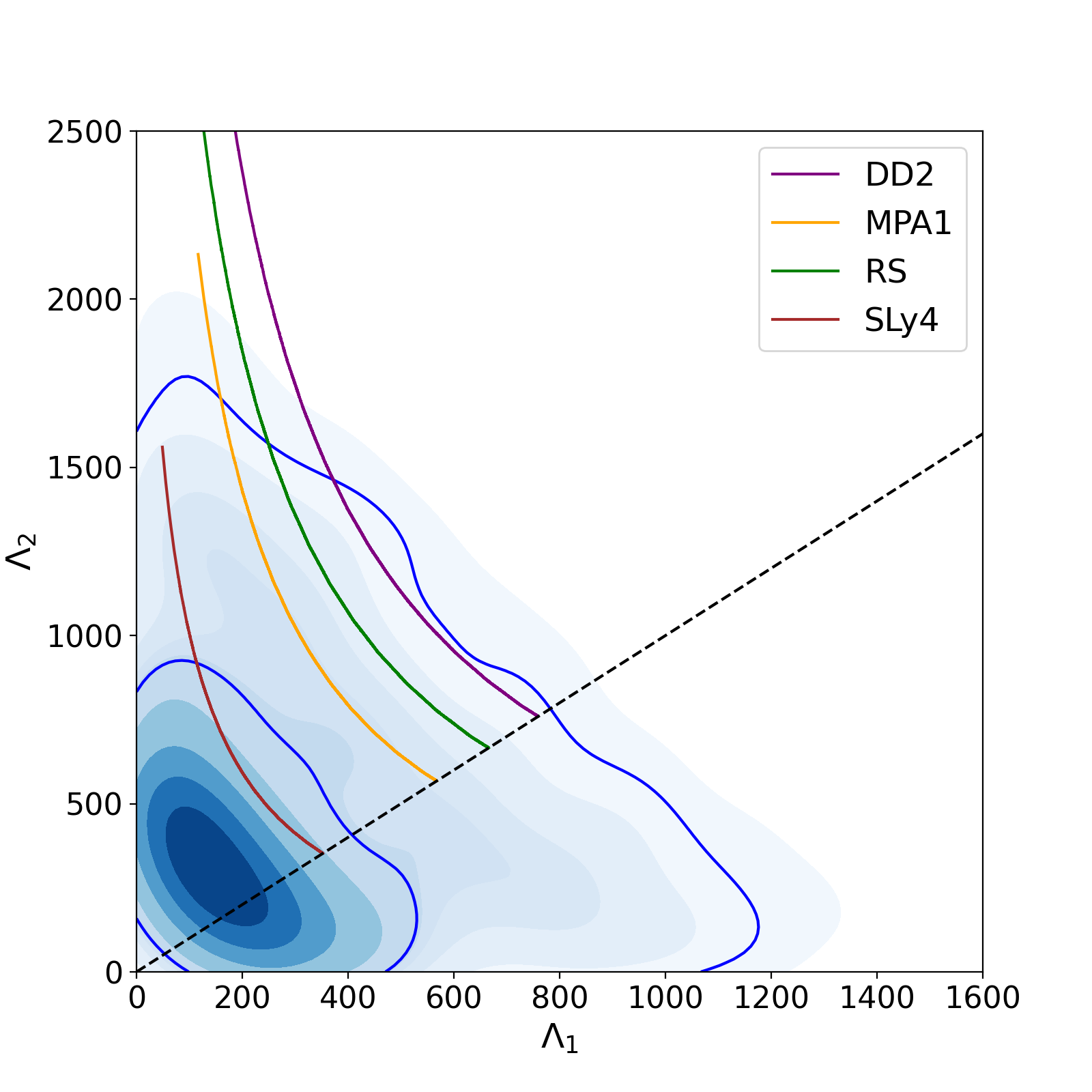}
    \caption{Posterior distribution (blue shading) of the dimensionless tidal deformabilities inferred from the GW170817 data, obtained drawing dimensionless spins uniformly up to 0.05 (low-spin prior) and by using the \texttt{IMRPhenomPv2\_NRTidal} waveform model \cite{Hannam_2014}. The solid blue lines indicate the 50\% and 90\% credible regions contours. The four colored curves represent the tidal parameters calculated for the EoS models we used for the injection sets. These are compatible with the constraint set by the 90\% credibility contour of the posterior distribution. The posterior distribution samples are taken from \cite{GW170817_prop, GW170817_data}.
}
    \label{fig:GW170817_constr}
\end{figure}

The procedure outlined above may also be applied outside the design search space to assess the performance of banks as one considers sources beyond the target space.  As stated in the Introduction, the focus of this work are tidal effects: we therefore consider a point-particle template bank and measure its effectualness against sets of injections with non-zero tidal deformabilities.  The sets differ in the EoS model assumed to determine the tidal deformability parameter, given the mass of an object in the binary source.
We choose four different EoSs: SLy4 \cite{Chabanat_1998, Douchin_2001}, MPA1 \cite{Muther_1987}, RS \cite{friedrich_1986, DANIELEWICZ200936, Gulminelli_2015}, DD2 \cite{Typel_2010, Grill_2014}. These are all compatible with the constraints imposed by the analysis of GW170817 \cite{GW170817_prop}, as shown in Figure \ref{fig:GW170817_constr}. Since tidal effects occur both in BNS and NSBH systems, for each EoS model we carry out the verification procedure with two sets of 10000 injections, one composed only by BNS signals and the other only by NSBH signals. In addition, for both type of sources, we also carry out for reference a bank verifier with point-like injections, i.e., setting tidal deformabilities to zero. All tests are performed with \texttt{pycbc\_make\_bank\_verifier\_workflow}, available in the PyCBC library \cite{PyCBC}.

The waveform approximant adopted to generate point-like injections and to model templates in the bank is \texttt{SEOBNRv5\_ROM} \cite{Buonanno_1999, Buonanno_2007, Bohe_2017, Pompili_2023}.
For consistence with this choice, injections with tidal effects are generated with the augmented version of the approximant, \texttt{SEOBNRv5\_ROM\_NRTidalv3} \cite{Abac_2024b}, for BNS systems and with \texttt{SEOBNRv4\_ROM\_NRTidalv2\_NSBH} \cite{Matas_2020} for NSBHs. 

The priors used to extract the values of the intrinsic parameters for each injection follow the structure of the parameter space nominally covered by the point-particle template bank. We therefore extract component masses from a uniform distribution over the intervals \( \left [1,2 \right] M_{\odot}\) and \( \left [2,50 \right] M_{\odot} \), for NSs and BHs, respectively;
further, we assume non-precessing binaries, with dimensionless spin magnitudes uniformly distributed over the range \( \left [-0.05, 0.05 \right] \). Tidal deformability is included using the dimensionless form of the parameter: \( \Lambda_{1,2} \equiv \lambda_{1,2}/{m_{1,2}}^5 \).
Specifically, \( \Lambda_{1} \) and \( \Lambda_{2} \) are calculated for each injection on the basis of the extracted values of the source-frame masses and of the EoS model adopted for the injection set in hand. In the case of NSBH systems, the tidal deformability of BHs is set identically to zero \cite{Chatziioannou_2020}. 

Finally, the extrinsic parameters are sampled as follows: the coalescence phase and the polarization angle are extracted uniformly in the range \( [0,2\pi] \); the inclination angle is taken within the interval \( [0,\pi] \) from a distribution uniform in its cosine; the sky-location is distributed uniformly over the sky, so that the declination is drawn within the interval \( [-\pi/2, \pi/2] \) from a distribution uniform in its cosine and the right ascension is drawn from a uniform distribution on \( [0, 2\pi] \); the luminosity distance is extracted from a uniform distribution from 100\,Mpc to 200\,Mpc.  
Both the starting frequency for the injected waveforms and the low-frequency cutoff used for the numerical computation of the match was fixed to \( 15 \) Hz, in agreement with the sensitivity curve adopted throughout this paper.

\begin{figure}[t!]
    \centering
    \includegraphics[width=1.0\linewidth]{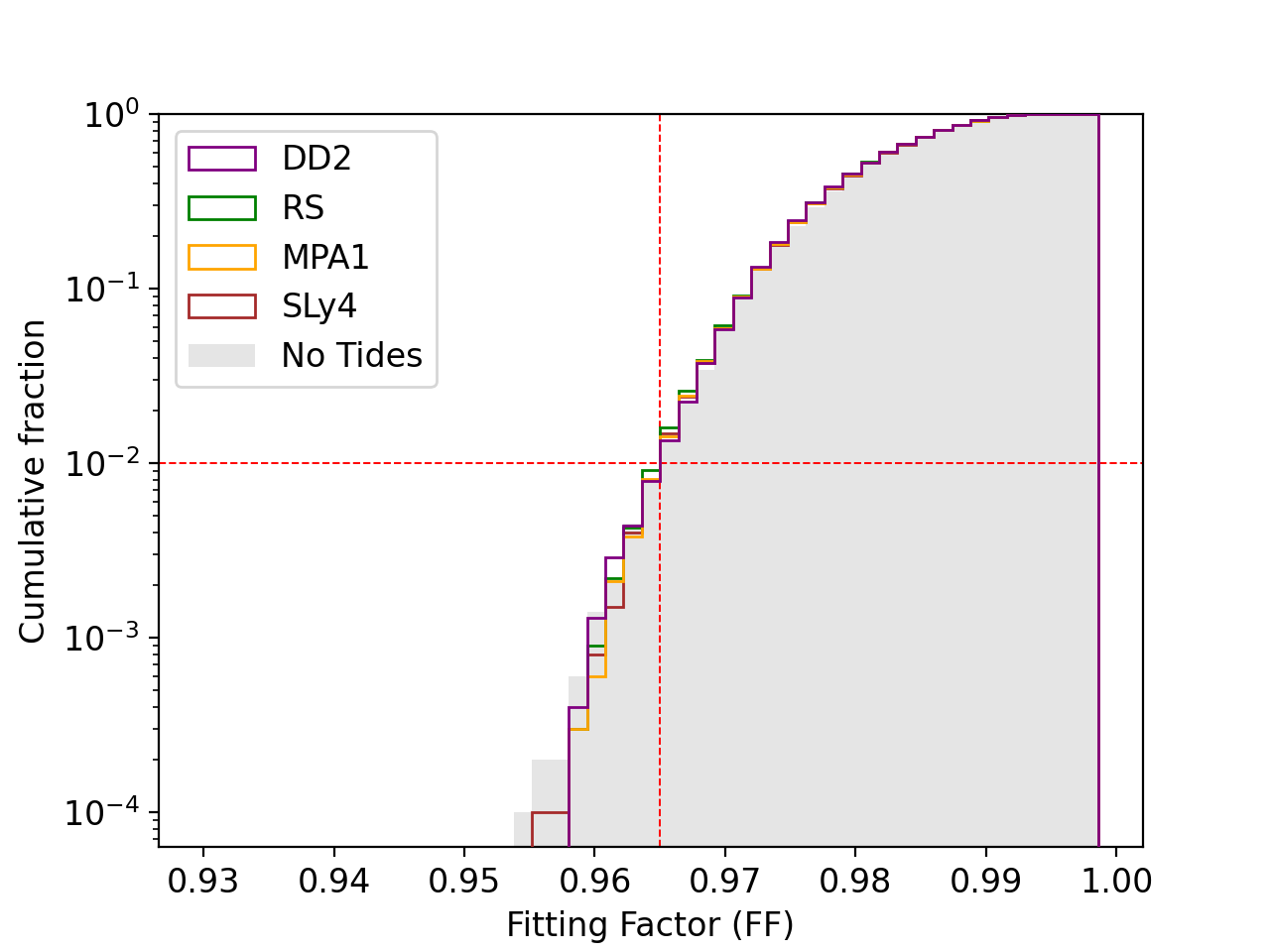}
    \includegraphics[width=1.0\linewidth]{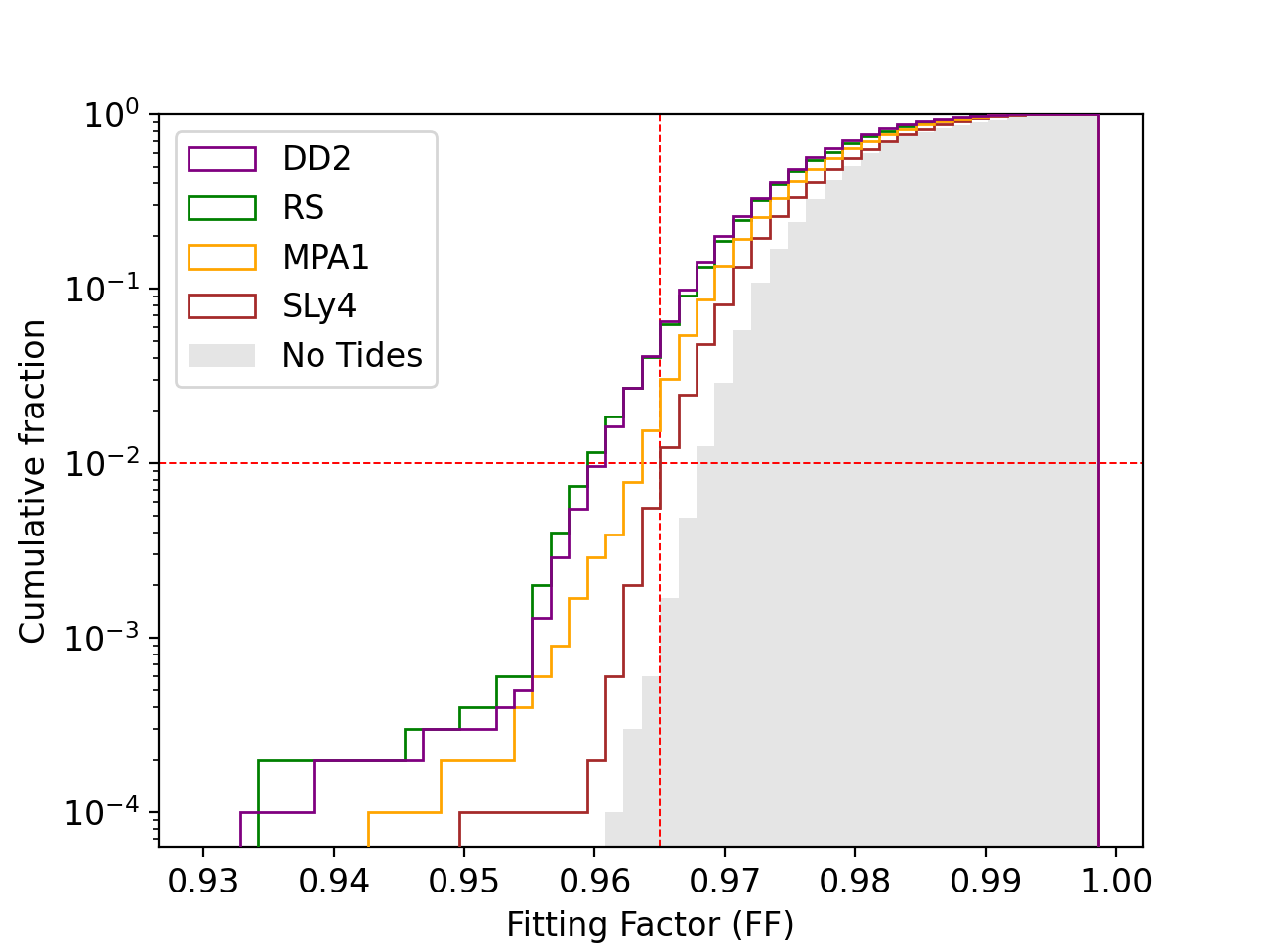}
    \par\vspace{2ex}
    \caption{Histograms of the cumulative distribution of the FF for the different injections sets against which we tested the performance of the point-particle template bank. Top panel: NSBH injections. Bottom panel: BNS injections. The vertical and horizontal red-dashed lines indicate the threshold of 1\% of injections at FF equal to 0.965, which is the minimum target match used when generating the template bank.
    }
    \label{fig:O4a_verifier}
\end{figure}

In Figure \ref{fig:O4a_verifier} we compare the cumulative FF distributions obtained for the various injection sets, for both BNS (top panel) and NSBH systems (bottom panel). It is evident that for NSBH systems the performance of the bank is not affected by the EoS chosen to calculate the tidal deformabilities of the injected waveforms; further it is equivalent to the performance against the point-like injections set. This is not surprising, because at the leading order tidal-induced effects result in a fifth order post-Newtonian (5PN) contribution term proportional to the mass-wighted tidal deformability parameter $\tilde{\Lambda}$ defined as \cite{Flanagan_2008, Blanchet_2024}:
\begin{equation}\label{eqn:lambda_tilde}
\tilde{\Lambda} = \frac{16}{13} \frac{(m_1 + 12 m_2)m_1^4 \Lambda_1 + (m_2 + 12 m_1)m_2^4 \Lambda_2}{(m_1+m_2)^5} \ .
\end{equation}
For NSBH systems, the term multiplying the tidal deformability of the BH vanishes and the total mass at the denominator suppresses the contribution: consequently, the tidal-induced 5PN correction term is generally smaller in the NSBH case, compared to the BNS case. Our first important conclusion is indeed the following:
\begin{itemize}
    \item \textit{for NSBH systems, we can exclude a loss of SNR beyond the one chosen in designing the template bank, even for banks that effectively treat these systems as point-particles, at least within a design representative of O4.}
\end{itemize}

\begin{table}[t!]
    \centering
    \caption{Performance of the point-particle template bank against the five injection sets representing BNS systems. The second and third columns report respectively the fraction of injections with FF lower than 0.965 (the template bank minimum match design value) and the minimum FF obtained over the whole set. \newline}
    \begin{tabular}{@{\hspace{0.2cm}}c@{\hspace{1cm}}c@{\hspace{1cm}}c@{\hspace{0.2cm}}}
        \addlinespace[0.5em]
        \toprule[1.3pt]
        \toprule[1.3pt]
        Injection set & FF $<$ 0.965 & Minimum FF \\ 
        \midrule[1.pt]
        DD2       & 4.14\% & 0.934 \\
        RS        & 4.09\% & 0.965 \\
        MPA1      & 1.54\% & 0.943 \\
        SLy4      & 0.56\%  & 0.951 \\
        No tides  & 0.06\%  & 0.962 \\
        \bottomrule[1.3pt]
        \bottomrule[1.3pt]
    \end{tabular}\label{tab:O4a_verifier}
\end{table}

The situation is radically different for BNS sources. Considering the histograms in the top panel of Figure \ref{fig:O4a_verifier}, we can indeed notice that the EoS-dependent injections have lower minimum FF values with respect to the point-like injections, and a larger fraction of injections with FF smaller than 0.965, which is the minimum match threshold with which the point-particle template bank under consideration is designed. These two quantities are reported in Table \ref{tab:O4a_verifier} for each set of BNS injections. We recall what we stated in Section \ref{sec:IIA}: for a bank to be validated as effectual, the fraction of injections returning a FF lower than the minimum match threshold must not be higher than 1\%. From Table \ref{tab:O4a_verifier}, we can see that the template bank is effectual against the point-like BNS injection set, as expected by design. However, for the injection sets with non-zero tidal deformabilities, the performance worsens as the EoS predicts more and more deformable neutron stars, or equivalently for less compressible NS matter. Against the injections generated with MPA1, RS and DD2 models, the bank cannot be considered effectual. 
Hence, our second set of conclusions is summarized as follows:
\begin{itemize}
    \item \textit{a point-particle template bank is not effectual when validated against waveforms representing BNS systems in which NS tidal deformability is not neglected, at least within a representative sensitivity of O4;}
    \item \textit{the effectualness of the bank decreases as the EoS model adopted sustains more deformable NSs.}
\end{itemize}
This confirms the findings of \cite{Cullen_2017}, that is, neglecting tidal-induced effects in the construction of the BNS region of a template bank causes a reduction in the recovered SNR and, therefore, in the sensitivity of the GW search.

The next section is dedicated to mitigating this short-coming.

\section{Building BNS template banks accounting for tidal deformability}\label{sec:III}
In this section we present our results on building a BNS template bank that accounts for tidal effects. Section \ref{sec:III.A} and \ref{sec:III.B} describe, respectively, the choice we made for the template placement algorithm and our new approach for incorporating tidal deformabilities into the target parameter space of the bank. Finally, Section \ref{sec:III.C} is dedicated to the discussion of the results we obtained in constructing and validating a BNS template bank of this kind.

\subsection{Template placement algorithms}\label{sec:III.A}
We already mentioned that the computational cost of building a template bank depends on the dimensionality and size of the parameter space. However, the algorithm adopted to spread templates across the parameter space plays a fundamental role as well \cite{Allen_2021}. Further, generating a bank arbitrarily increasing the number of templates is not a viable option as it implies higher and higher computational costs for the search, and it increases the false alarm rate threshold to separate signals from noise, as more and more noise transients will ``ring'' with available templates. The placement algorithm must aim at reducing as much as possible the number of templates while keeping under control signal losses due to the discretization of the parameter space. 

Different template placement methods have been developed over the years. The effectiveness of one method over others is not general; rather, it depends on the characteristics of the parameter space to be covered. We can identify two broad categories: geometric \cite{Babak_2006, Brown_2012, Harry_2014, Hanna_2022, Phukon:2024amh} and stochastic \cite{Babak_2008, Ajith_2014, Capanno_2016, Fehrmann_2014} placement.

\noindent{Geometric placement methods} are based on lattice strategies to place templates. An approximate analytic expression for the match between template waveforms is defined and adopted as a metric on the parameter space manifold. Templates are then spread on the manifold using a reduced-dimension lattice of points, distanced according to the design minimum match threshold \cite{Owen_1996, Owen_1999, Cokelaer_2007}. 
\citet{Harry_Lundgren} demonstrated that geometric algorithms become unreliable in predicting waveform matches when considering PN term variations beyond the 4PN order, unless the match is approximated at sufficiently high orders, which would however make the search computationally prohibitive \cite{Sharma:2023djw}.
Hence, geometric placement algorithms are not suitable for constructing BNS template banks that include tidal deformations effects, such as those we aim to build. Following the recommendation of \citet{Harry_Lundgren}, we adopted a stochastic placement method. This consists in using additive brute-force algorithms that are indeed more flexible and efficient when handling arbitrary parameter spaces, because they do not require the definition of an explicit metric and a lattice-based strategy.
Specifically, we made use of \texttt{pycbc\_brute\_bank} from the \texttt{PyCBC} library \cite{PyCBC}. A comprehensive description of this tool is provided in \cite{kacanja_2024}; here we just summarize its key features.  
Putative templates are randomly sampled from prior distributions representing the target parameter space. Then, their fitting factor against all the previously retained templates is calculated, if it is below the design minimum match threshold the proposal is added to the bank, otherwise it is rejected. To optimize coverage in less densely populated regions of the parameter space, the algorithm switches to drawing proposals based on a Gaussian Kernel Density Estimate \cite{Parzen_1962, Falxa_2023} when this is more efficient than a uniform extraction from the priors. A number of design choices were made by the authors to reduce the computational burden and speed up match calculations during the process of template acceptance. Primarily, all waveforms are generated with a reduced order frequency \cite{Ajith_2014, Capanno_2016}, and the triangle inequality is used to efficiently estimate matches based on previously computed ones.

\subsection{Tidal deformability prior}\label{sec:III.B}

\begin{figure}
    \centering
    \includegraphics[width=1\linewidth]{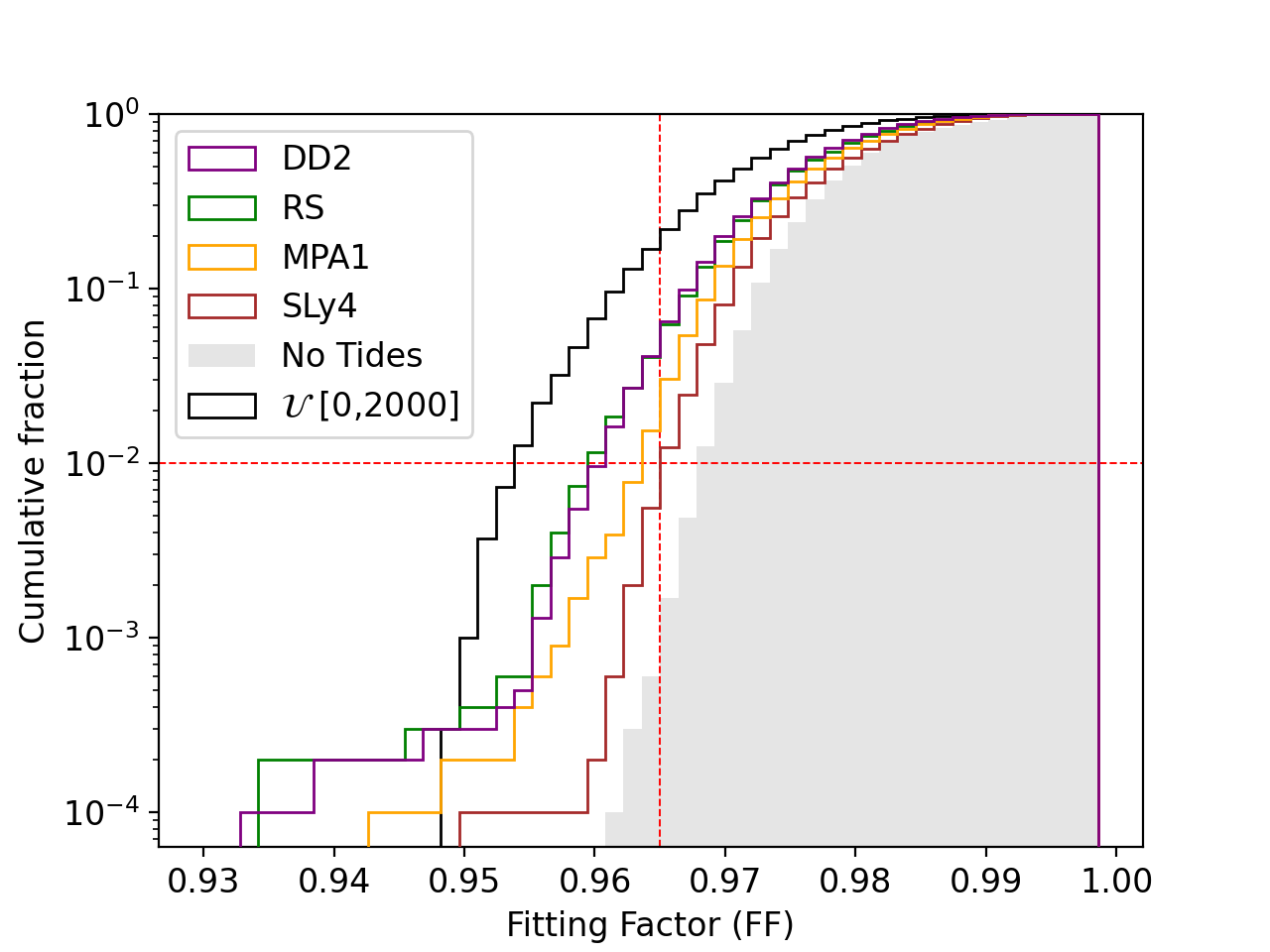}
    \caption{Same plot as in Figure \ref{fig:O4a_verifier} with the addition of the histogram relative to the set of injections with dimensionless tidal deformability values randomly extracted between 0 and 2000 (black line).}
    \label{fig:O4a_verifier_2}
\end{figure}

\begin{table}[t!]
    \centering
    \caption{Same table as in Table \ref{tab:O4a_verifier} with the addition of the results relative to the set of injections with dimensionless tidal deformabilities values randomly extracted between 0 and 2000 (bottom row).\newline}
    \begin{tabular}{@{\hspace{0.2cm}}c@{\hspace{1cm}}c@{\hspace{1cm}}c@{\hspace{0.2cm}}}
        \addlinespace[0.5em]
        \toprule[1.3pt]
        \toprule[1.3pt]
        Injection set & FF $<$ 0.965 & Minimum FF \\ 
        \midrule[1.pt]
        DD2       & 4.14\% & 0.934 \\
        RS        & 4.09\% & 0.965 \\
        MPA1      & 1.54\% & 0.943 \\
        SLy4      & 0.56\%  & 0.951 \\
        No tides  & 0.06\%  & 0.962 \\
        \(\mathcal{U}[0,2000]\) & 16.9\% & 0.950 \\
        \bottomrule[1.3pt]
        \bottomrule[1.3pt]
    \end{tabular}\label{tab:O4a_verifier_2}
\end{table}

In order to include the dimensionless tidal deformabilities \( ( \Lambda_1, \Lambda_2 ) \) in the target parameter space, we begin by considering the proposal of \citet{Harry_Lundgren} who suggested extracting the values of both parameters from a uniform distribution over the range \( [0, 2000]\).
We test this prescription by carrying out a new effectualness test on the point-particle template bank against an set of injections representing a proxy of the prior proposed in \cite{Harry_Lundgren} for the template bank construction. In this new set, each injection has values of \( \Lambda_1 \) and \( \Lambda_2 \) uniformly drawn in the range [0, 2000], rather than calculated from \(m_1\) and \(m_2\) according to a specific EoS model. Figure \ref{fig:O4a_verifier_2} and Table \ref{fig:O4a_verifier_2} summarize the results of this new bank verifier and compare them to those already presented in \ref{sec:IIB}. The performance of the point-particle template bank is worse than that witnessed for the injections built with the DD2 EoS model, which is representative of the upper boundary of the GW170817 constraints (cfr.~Figure \ref{fig:GW170817_constr}). Specifically, the fraction of injections returning an FF below 0.965 is $\sim 4$ times the corresponding value for the DD2 injection set. This result indicates that the flat prior would overestimate the need for the bank to cover tidal effects. Applying such prior for the extraction of $\Lambda$'s in the construction of a template bank, would therefore lead to overcoverage, that is, a number of templates larger than necessary.  As a consequence, the search computing costs and false alarm rate threshold would grow excessively. 

This outcome and the need to include tidal effects in BNS template banks demonstrated by Figure \ref{fig:O4a_verifier} motivate us to formulate a physically-informed approach to set up the prior needed to extract the tidal deformability values.  Our prescription is formulated as follows:

\noindent\textit{for each proposal, after the component masses \( ( m_1, m_2 ) \) are extracted, the associated \( (\Lambda_1 \), \( \Lambda_2) \) are drawn from a uniform distribution between 0 and the value computed with the DD2 EoS model for the relevant component mass.}

\noindent In this approach, we again consider the DD2 model as representative of the upper limit of the GW170817 constraint, in an attempt to accurately fold in our uncertainty on the NS EoS. We can express our proposal formally as: \textit{the extraction of \( ( m_1, m_2 ) \) is followed by the extraction of $\Lambda_i\in\mathcal{U}[0, \Lambda_{\rm DD2}(m_i)]$ for i=\{1, 2\}.}

We stress that, by design, our method improves as the constraints on the EoS become tighter.

\subsection{Results}\label{sec:III.C}

We now consider three distinct BNS template banks: one in which tidal deformability values are extracted using our physically-informed prior; a second one in which tidal deformability values are extracted from the uniform prior proposed in \cite{Harry_Lundgren}; and a third one in which they are set to zero, i.e., a point-particle BNS template bank. The aim is to compare the number of additional templates required to include tidal deformabilities in the parameter space. 
The banks are all generated with the same priors for masses and spins; namely, masses are extracted from a uniform prior over the interval \( \left [1,2 \right] M_{\odot}\), while dimensionless aligned-spins are sampled over the range \( \left [-0.05, 0.05 \right] \). 
Similarly to the tests described in Section \ref{sec:IIB}, the approximant adopted for templates of the first two banks with tidal effects is \texttt{SEOBNRv5\_ROM\_NRTidalv3}, while its tidal-less version \texttt{SEOBNRv5\_ROM} is used for the third, point-particle bank.
We utilized again the LIGO O4 design sensitivity as noise-power spectral density, a lower frequency cutoff of 15\,Hz and a target fitting factor of 0.97.
The number of templates in each of the banks is reported in Table \ref{tab:BNS_banks}.
We can conclude that: 
\begin{itemize}
    \item \textit{The BNS template bank built with our physically-informed prior for the NS tidal deformabilities has \textbf{8.23\%} additional templates with respect to the point-like bank built with the same choice of parameter space, but neglecting tidal deformability. This result represents a substantial reduction with respect to the bank augmented with the fixed-range flat prior, which has \textbf{32.9\%} extra templates than the point-like bank.}
\end{itemize}

\begin{table}[t!]
    \centering
    \caption{Number of templates for the BNS template banks described in Section \ref{sec:III.C}.}
    \begin{tabular}{@{\hspace{0.2cm}}c@{\hspace{1cm}}c@{\hspace{0.2cm}}}
        \addlinespace[0.5em]
        \toprule[1.3pt]
        \toprule[1.3pt]
        Template bank & Number of templates \\
        \midrule[1.pt]
        $\Lambda_i \in \mathcal{U}[0, \Lambda_{{\rm DD2}}(m_i)]$ & 51210 \\
        $\Lambda_i \in \mathcal{U}[0, 2000]$ & 62898 \\
        $\Lambda_i = 0$ & 47315 \\
        \bottomrule[1.3pt]
        \bottomrule[1.3pt]
    \end{tabular}\label{tab:BNS_banks}
\end{table}

\begin{figure}
    \centering
    \includegraphics[width=1\linewidth]{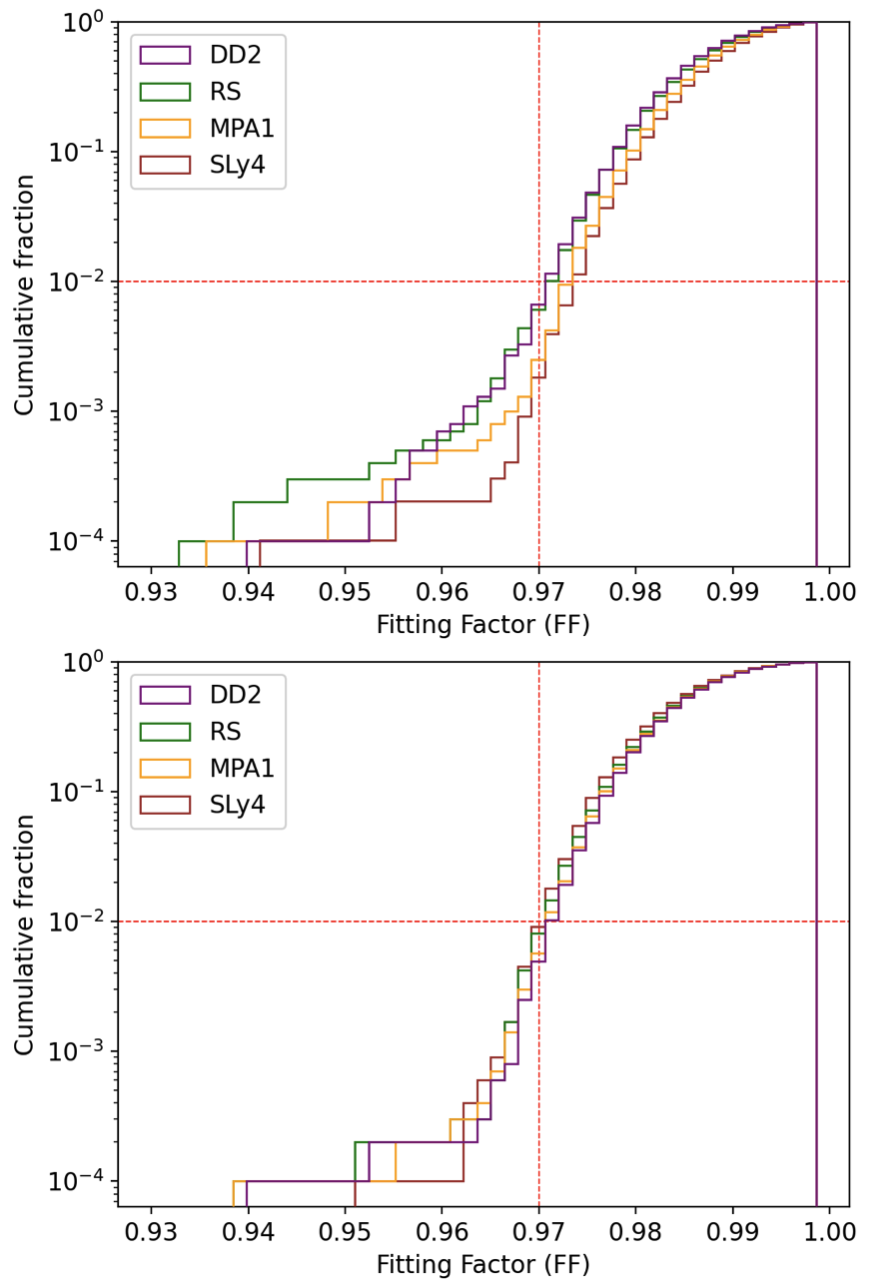}
    \caption{Histograms of the cumulative distribution of the FF for the four sets of injections used to test the performance of the BNS template bank generated with our physically-informed prior for extracting tidal deformability values (top panel), and the BNS template bank generated with the fixed-range flat prior for extracting tidal deformability values (bottom panel). The red-dashed vertical and horizontal lines indicate the threshold of 1\% of injections and FF equal to 0.97, respectively.}
    \label{fig:BNS_verifiers}
\end{figure}

As a final step, we assess the effectualness of the BNS template bank constructed adopting the prior proposed in this work. To this purpose, we performed on this bank effectualness tests identical to those described in Section \ref{sec:IIB} for the generic CBC point-particle template bank. For completeness, we carried out such verifiers also on the template bank built with the fixed-range prior. The results of these tests are reported in Figure \ref{fig:BNS_verifiers} and Table \ref{tab:BNS_verifiers}. We observe that:
\begin{itemize}
\item \textit{Unlike the point-particle template bank considered in Section \ref{sec:II}, the template bank generated with our new prior is effectual against all the injection sets with tides.}

\item \textit{The performance of the two template banks are similar against all sets of injections. However, as pointed out before, the number of templates obtained with our prior is considerably lower.}
\end{itemize}

\begin{table}[t!]
    \centering
    \caption{Fraction of injections with FF below 0.97 for the different sets of injections (identified by the first label in each row) against which we tested the performance of the two tidal template banks considered in this work (identified by the prior ranges in the last two columns).}
     \begin{tabular}{@{\hspace{0.2cm}}c@{\hspace{1cm}}c@{\hspace{1cm}}c@{\hspace{0.2cm}}}
        \addlinespace[0.5em]
        \toprule[1.3pt]
        \toprule[1.3pt]
         & \multicolumn{2}{c}{$\Lambda_i$ uniform prior range} \\
        Injection set & $[0, \Lambda_{{\rm DD2}}(m_i)]$ & $[0, 2000]$ \\
        \midrule[1.pt]
        DD2  & 0.52\% & 0.34\% \\
        RS   & 0.55\% & 0.64\% \\
        MPA1 & 0.21\% & 0.46\% \\
        SLy4 & 0.16\% & 0.70\% \\
        \bottomrule[1.3pt]
        \bottomrule[1.3pt]
    \end{tabular}\label{tab:BNS_verifiers}
\end{table}

\section{Conclusions}
In this paper we presented a new technique to account for the effect of NS tidal deformations when constructing template banks for matched-filtering searches of CBC signals. 
We argued that ignoring these effects, as is currently done, can lead to signal losses.  We did so testing the effectualness of a point-particle template bank built with a sensitivity curve representative of O4 against sets of BNS and NSBH simulated signals with non-zero tidal deformabilities.
We considered various sets of injections by selecting a different EoS model to compute the NS tidal deformability value corresponding to a specific mass.  All EoSs considered are compatible with constraints provided by the GW170817 observation. 
We found no evidence of signal losses in the case of NSBH systems, while for BNS systems we observed a non-negligible reduction in sensitivity of the bank, which increased as we assumed stiffer NS EoS models.

We then addressed the need to cover tidal effects with template banks, while minimizing the required number of additional templates as much as possible.  
As prescribed in \cite{Harry_Lundgren}, we adopted a stochastic template placement algorithm, specifically \texttt{pycbc\_brute\_bank} \cite{kacanja_2024} from the \texttt{PyCBC} library \cite{PyCBC}. Our approach consists in extracting the NS tidal deformability value corresponding to the extracted mass value of a given template proposal from a uniform distribution between 0 and the value computed with the DD2 EoS model for that specific mass. The DD2 model was chosen based on the GW170817 constraints (see Figure \ref{fig:GW170817_constr}).

We compared a BNS bank built with our proposal to a point-particle bank covering the same parameter space, with the exception of tidal deformabilities, and one build with the prescription suggested in \cite{Harry_Lundgren} of extracting (dimensionless) tidal deformabilities uniformly between 0 and 2000.
The effectualness tests of these three banks show that the two banks with coverage of tidal deformabilities are effectual against simulated signals with tides, confirming the validity of our approach in covering tidal effects. At the same time, the bank obtained with the approach proposed in this paper has only 8.23\% additional templates with respect to the equivalent point-like bank, as opposed to the 32.9\% additional templates yielded by the prescription of \cite{Harry_Lundgren}.

We therefore achieved coverage of tidal effects, and hence a reduction of potential BNS signal losses, while preventing a substantial increase of the size of the template bank.

\begin{acknowledgments}
We wish to thank Ian Harry for useful discussions. We acknowledge support from the ICSC - Centro Nazionale di Ricerca in High Performance Computing, Big Data and Quantum Computing, funded by the European Union - NextGenerationEU and support from the Italian Ministry of University and Research (MUR) Progetti di ricerca di Rilevante Interesse Nazionale (PRIN) Bando 2022 - grant 20228TLHPE - CUP I53D23000630006.

\end{acknowledgments}

\bibliography{apssamp}

\providecommand{\noopsort}[1]{}\providecommand{\singleletter}[1]{#1}%
\begin{thebibliography}{72}%
\makeatletter
\providecommand \@ifxundefined [1]{%
 \@ifx{#1\undefined}
}%
\providecommand \@ifnum [1]{%
 \ifnum #1\expandafter \@firstoftwo
 \else \expandafter \@secondoftwo
 \fi
}%
\providecommand \@ifx [1]{%
 \ifx #1\expandafter \@firstoftwo
 \else \expandafter \@secondoftwo
 \fi
}%
\providecommand \natexlab [1]{#1}%
\providecommand \enquote  [1]{``#1''}%
\providecommand \bibnamefont  [1]{#1}%
\providecommand \bibfnamefont [1]{#1}%
\providecommand \citenamefont [1]{#1}%
\providecommand \href@noop [0]{\@secondoftwo}%
\providecommand \href [0]{\begingroup \@sanitize@url \@href}%
\providecommand \@href[1]{\@@startlink{#1}\@@href}%
\providecommand \@@href[1]{\endgroup#1\@@endlink}%
\providecommand \@sanitize@url [0]{\catcode `\\12\catcode `\$12\catcode `\&12\catcode `\#12\catcode `\^12\catcode `\_12\catcode `\%12\relax}%
\providecommand \@@startlink[1]{}%
\providecommand \@@endlink[0]{}%
\providecommand \url  [0]{\begingroup\@sanitize@url \@url }%
\providecommand \@url [1]{\endgroup\@href {#1}{\urlprefix }}%
\providecommand \urlprefix  [0]{URL }%
\providecommand \Eprint [0]{\href }%
\providecommand \doibase [0]{https://doi.org/}%
\providecommand \selectlanguage [0]{\@gobble}%
\providecommand \bibinfo  [0]{\@secondoftwo}%
\providecommand \bibfield  [0]{\@secondoftwo}%
\providecommand \translation [1]{[#1]}%
\providecommand \BibitemOpen [0]{}%
\providecommand \bibitemStop [0]{}%
\providecommand \bibitemNoStop [0]{.\EOS\space}%
\providecommand \EOS [0]{\spacefactor3000\relax}%
\providecommand \BibitemShut  [1]{\csname bibitem#1\endcsname}%
\let\auto@bib@innerbib\@empty
\bibitem [{\citenamefont {Abbott}\ \emph {et~al.}(2016)\citenamefont {Abbott} \emph {et~al.}}]{GW150914}%
  \BibitemOpen
  \bibfield  {author} {\bibinfo {author} {\bibfnamefont {B.}~\bibnamefont {Abbott}} \emph {et~al.} (\bibinfo {collaboration} {LIGO Scientific Collaboration and Virgo Collaboration}),\ }\bibfield  {title} {\bibinfo {title} {{Observation of Gravitational Waves from a Binary Black Hole Merger}},\ }\href {https://doi.org/10.1103/PhysRevLett.116.061102} {\bibfield  {journal} {\bibinfo  {journal} {Phys. Rev. Lett.}\ }\textbf {\bibinfo {volume} {116}},\ \bibinfo {pages} {061102} (\bibinfo {year} {2016})}\BibitemShut {NoStop}%
\bibitem [{\citenamefont {Abac}\ \emph {et~al.}(2025{\natexlab{a}})\citenamefont {Abac} \emph {et~al.}}]{GWTC_4_intro}%
  \BibitemOpen
  \bibfield  {author} {\bibinfo {author} {\bibfnamefont {A.~G.}\ \bibnamefont {Abac}} \emph {et~al.} (\bibinfo {collaboration} {LIGO Scientific, VIRGO, KAGRA}),\ }\href@noop {} {\bibinfo {title} {{GWTC-4.0: An Introduction to Version 4.0 of the Gravitational-Wave Transient Catalog}}} (\bibinfo {year} {2025}{\natexlab{a}}),\ \Eprint {https://arxiv.org/abs/2508.18080} {arXiv:2508.18080 [gr-qc]} \BibitemShut {NoStop}%
\bibitem [{\citenamefont {Abac}\ \emph {et~al.}(2025{\natexlab{b}})\citenamefont {Abac} \emph {et~al.}}]{GWTC_4_methods}%
  \BibitemOpen
  \bibfield  {author} {\bibinfo {author} {\bibfnamefont {A.~G.}\ \bibnamefont {Abac}} \emph {et~al.} (\bibinfo {collaboration} {LIGO Scientific Collaboration, Virgo Collaboration, and KAGRA Collaboration}),\ }\href {https://arxiv.org/abs/2508.18081} {\bibinfo {title} {Gwtc-4.0: Methods for identifying and characterizing gravitational-wave transients}} (\bibinfo {year} {2025}{\natexlab{b}}),\ \Eprint {https://arxiv.org/abs/2508.18081} {arXiv:2508.18081 [gr-qc]} \BibitemShut {NoStop}%
\bibitem [{\citenamefont {Abac}\ \emph {et~al.}(2025{\natexlab{c}})\citenamefont {Abac} \emph {et~al.}}]{GWTC_4_results}%
  \BibitemOpen
  \bibfield  {author} {\bibinfo {author} {\bibfnamefont {A.~G.}\ \bibnamefont {Abac}} \emph {et~al.} (\bibinfo {collaboration} {LIGO Scientific, VIRGO, KAGRA}),\ }\href@noop {} {\bibinfo {title} {{GWTC-4.0: Updating the Gravitational-Wave Transient Catalog with Observations from the First Part of the Fourth LIGO-Virgo-KAGRA Observing Run}}} (\bibinfo {year} {2025}{\natexlab{c}}),\ \Eprint {https://arxiv.org/abs/2508.18082} {arXiv:2508.18082 [gr-qc]} \BibitemShut {NoStop}%
\bibitem [{\citenamefont {Abac}\ \emph {et~al.}(2025{\natexlab{d}})\citenamefont {Abac} \emph {et~al.}}]{GWTC_4_GWOSC}%
  \BibitemOpen
  \bibfield  {author} {\bibinfo {author} {\bibfnamefont {A.~G.}\ \bibnamefont {Abac}} \emph {et~al.} (\bibinfo {collaboration} {LIGO Scientific, VIRGO, KAGRA}),\ }\href@noop {} {\bibinfo {title} {{Open Data from LIGO, Virgo, and KAGRA through the First Part of the Fourth Observing Run}}} (\bibinfo {year} {2025}{\natexlab{d}}),\ \Eprint {https://arxiv.org/abs/2508.18079} {arXiv:2508.18079 [gr-qc]} \BibitemShut {NoStop}%
\bibitem [{\citenamefont {Aasi}\ \emph {et~al.}(2015)\citenamefont {Aasi} \emph {et~al.}}]{Aasi_2015}%
  \BibitemOpen
  \bibfield  {author} {\bibinfo {author} {\bibfnamefont {J.}~\bibnamefont {Aasi}} \emph {et~al.} (\bibinfo {collaboration} {LIGO Scientific Collaboration}),\ }\bibfield  {title} {\bibinfo {title} {{Advanced LIGO}},\ }\href {https://doi.org/10.1088/0264-9381/32/7/074001} {\bibfield  {journal} {\bibinfo  {journal} {Classical and Quantum Gravity}\ }\textbf {\bibinfo {volume} {32}},\ \bibinfo {pages} {074001} (\bibinfo {year} {2015})}\BibitemShut {NoStop}%
\bibitem [{\citenamefont {Acernese}\ \emph {et~al.}(2014)\citenamefont {Acernese} \emph {et~al.}}]{Acernese_2015}%
  \BibitemOpen
  \bibfield  {author} {\bibinfo {author} {\bibfnamefont {F.}~\bibnamefont {Acernese}} \emph {et~al.} (\bibinfo {collaboration} {Virgo Collaboration}),\ }\bibfield  {title} {\bibinfo {title} {{Advanced Virgo: a second-generation interferometric gravitational wave detector}},\ }\href {https://doi.org/10.1088/0264-9381/32/2/024001} {\bibfield  {journal} {\bibinfo  {journal} {Classical and Quantum Gravity}\ }\textbf {\bibinfo {volume} {32}},\ \bibinfo {pages} {024001} (\bibinfo {year} {2014})}\BibitemShut {NoStop}%
\bibitem [{\citenamefont {Akutsu}\ \emph {et~al.}(2021)\citenamefont {Akutsu} \emph {et~al.}}]{KAGRA:2020tym}%
  \BibitemOpen
  \bibfield  {author} {\bibinfo {author} {\bibfnamefont {T.}~\bibnamefont {Akutsu}} \emph {et~al.} (\bibinfo {collaboration} {KAGRA}),\ }\bibfield  {title} {\bibinfo {title} {{Overview of KAGRA: Detector design and construction history}},\ }\href {https://doi.org/10.1093/ptep/ptaa125} {\bibfield  {journal} {\bibinfo  {journal} {PTEP}\ }\textbf {\bibinfo {volume} {2021}},\ \bibinfo {pages} {05A101} (\bibinfo {year} {2021})},\ \Eprint {https://arxiv.org/abs/2005.05574} {arXiv:2005.05574 [physics.ins-det]} \BibitemShut {NoStop}%
\bibitem [{\citenamefont {Abac}\ \emph {et~al.}(2025{\natexlab{e}})\citenamefont {Abac} \emph {et~al.}}]{GW250114_discovery}%
  \BibitemOpen
  \bibfield  {author} {\bibinfo {author} {\bibfnamefont {A.~G.}\ \bibnamefont {Abac}} \emph {et~al.} (\bibinfo {collaboration} {KAGRA, Virgo, LIGO Scientific}),\ }\bibfield  {title} {\bibinfo {title} {{GW250114: Testing Hawking{\textquoteright}s Area Law and the Kerr Nature of Black Holes}},\ }\href {https://doi.org/10.1103/kw5g-d732} {\bibfield  {journal} {\bibinfo  {journal} {Phys. Rev. Lett.}\ }\textbf {\bibinfo {volume} {135}},\ \bibinfo {pages} {111403} (\bibinfo {year} {2025}{\natexlab{e}})},\ \Eprint {https://arxiv.org/abs/2509.08054} {arXiv:2509.08054 [gr-qc]} \BibitemShut {NoStop}%
\bibitem [{Note1()}]{Note1}%
  \BibitemOpen
  \bibinfo {note} {\protect \url {https://gracedb.ligo.org}.}\BibitemShut {Stop}%
\bibitem [{\citenamefont {Abbott}\ \emph {et~al.}(2017{\natexlab{a}})\citenamefont {Abbott} \emph {et~al.}}]{GW170817}%
  \BibitemOpen
  \bibfield  {author} {\bibinfo {author} {\bibfnamefont {B.}~\bibnamefont {Abbott}} \emph {et~al.} (\bibinfo {collaboration} {LIGO Scientific Collaboration and Virgo Collaboration}),\ }\bibfield  {title} {\bibinfo {title} {{GW170817: Observation of Gravitational Waves from a Binary Neutron Star Inspiral}},\ }\href {https://doi.org/10.1103/PhysRevLett.119.161101} {\bibfield  {journal} {\bibinfo  {journal} {Phys. Rev. Lett.}\ }\textbf {\bibinfo {volume} {119}},\ \bibinfo {pages} {161101} (\bibinfo {year} {2017}{\natexlab{a}})}\BibitemShut {NoStop}%
\bibitem [{\citenamefont {Abbott}\ \emph {et~al.}(2020{\natexlab{a}})\citenamefont {Abbott} \emph {et~al.}}]{Abbott_2020}%
  \BibitemOpen
  \bibfield  {author} {\bibinfo {author} {\bibfnamefont {B.}~\bibnamefont {Abbott}} \emph {et~al.} (\bibinfo {collaboration} {LIGO Scientific Collaboration and Virgo Collaboration}),\ }\bibfield  {title} {\bibinfo {title} {{GW190425: Observation of a Compact Binary Coalescence with Total Mass \(\sim\) 3.4 $M_\odot$}},\ }\href {https://doi.org/10.3847/2041-8213/ab75f5} {\bibfield  {journal} {\bibinfo  {journal} {The Astrophysical Journal Letters}\ }\textbf {\bibinfo {volume} {892}},\ \bibinfo {pages} {L3} (\bibinfo {year} {2020}{\natexlab{a}})}\BibitemShut {NoStop}%
\bibitem [{\citenamefont {Abbott}\ \emph {et~al.}(2021)\citenamefont {Abbott} \emph {et~al.}}]{Abbott_2021}%
  \BibitemOpen
  \bibfield  {author} {\bibinfo {author} {\bibfnamefont {R.}~\bibnamefont {Abbott}} \emph {et~al.} (\bibinfo {collaboration} {LIGO Scientific Collaboration, Virgo Collaboration, and KAGRA Collaboration}),\ }\bibfield  {title} {\bibinfo {title} {{Observation of Gravitational Waves from Two Neutron Star–Black Hole Coalescences}},\ }\href {https://doi.org/10.3847/2041-8213/ac082e} {\bibfield  {journal} {\bibinfo  {journal} {The Astrophysical Journal Letters}\ }\textbf {\bibinfo {volume} {915}},\ \bibinfo {pages} {L5} (\bibinfo {year} {2021})}\BibitemShut {NoStop}%
\bibitem [{\citenamefont {Abac}\ \emph {et~al.}(2024{\natexlab{a}})\citenamefont {Abac} \emph {et~al.}}]{Abac_2024}%
  \BibitemOpen
  \bibfield  {author} {\bibinfo {author} {\bibfnamefont {A.}~\bibnamefont {Abac}} \emph {et~al.} (\bibinfo {collaboration} {LIGO Scientific Collaboration, Virgo Collaboration, and KAGRA Collaboration}),\ }\bibfield  {title} {\bibinfo {title} {{Observation of Gravitational Waves from the Coalescence of a 2.5–4.5 $M_\odot$ Compact Object and a Neutron Star}},\ }\href {https://doi.org/10.3847/2041-8213/ad5beb} {\bibfield  {journal} {\bibinfo  {journal} {The Astrophysical Journal Letters}\ }\textbf {\bibinfo {volume} {970}},\ \bibinfo {pages} {L34} (\bibinfo {year} {2024}{\natexlab{a}})}\BibitemShut {NoStop}%
\bibitem [{\citenamefont {Abbott}\ \emph {et~al.}(2019{\natexlab{a}})\citenamefont {Abbott} \emph {et~al.}}]{GW170817_prop}%
  \BibitemOpen
  \bibfield  {author} {\bibinfo {author} {\bibfnamefont {B.}~\bibnamefont {Abbott}} \emph {et~al.} (\bibinfo {collaboration} {LIGO Scientific Collaboration and Virgo Collaboration}),\ }\bibfield  {title} {\bibinfo {title} {{Properties of the Binary Neutron Star Merger GW170817}},\ }\href {https://doi.org/10.1103/PhysRevX.9.011001} {\bibfield  {journal} {\bibinfo  {journal} {Phys. Rev. X}\ }\textbf {\bibinfo {volume} {9}},\ \bibinfo {pages} {011001} (\bibinfo {year} {2019}{\natexlab{a}})}\BibitemShut {NoStop}%
\bibitem [{\citenamefont {Abbott}\ \emph {et~al.}(2017{\natexlab{b}})\citenamefont {Abbott} \emph {et~al.}}]{GW170817_multi}%
  \BibitemOpen
  \bibfield  {author} {\bibinfo {author} {\bibfnamefont {B.~P.}\ \bibnamefont {Abbott}} \emph {et~al.} (\bibinfo {collaboration} {LIGO Scientific Collaboration and Virgo Collaboration, Fermi GBM, INTEGRAL, IceCube Collaboration, AstroSat Cadmium Zinc Telluride Imager Team, IPN Collaboration, The Insight-HXMT Collaboration, ANTARES Collaboration, The Swift Collaboration, AGILE Team, The 1M2H Team, The Dark Energy Camera GW-EM Collaboration and the DES Collaboration, The DLT40 Collaboration, GRAWITA: GRAvitational Wave Inaf TeAm, The Fermi Large Area Telescope Collaboration, ATCA: Australia Telescope Compact Array, ASKAP: Australian SKA Pathfinder, Las Cumbres Observatory Group, OzGrav, DWF (Deeper, Wider, Faster Program), AST3, and CAASTRO Collaborations, The VINROUGE Collaboration, MASTER Collaboration, J-GEM, GROWTH, JAGWAR, CaltechNRAO, TTU-NRAO, and NuSTAR Collaborations, Pan-STARRS, The MAXI Team, TZAC Consortium, KU Collaboration, Nordic Optical Telescope, ePESSTO, GROND, Texas Tech University, SALT
  Group, TOROS: Transient Robotic Observatory of the South Collaboration, The BOOTES Collaboration, MWA: Murchison Widefield Array, The CALET Collaboration, IKI-GW Follow-up Collaboration, H.E.S.S. Collaboration, LOFAR Collaboration, LWA: Long Wavelength Array, HAWC Collaboration, The Pierre Auger Collaboration, ALMA Collaboration, Euro VLBI Team, Pi of the Sky Collaboration, The Chandra Team at McGill University, DFN: Desert Fireball Network, ATLAS, High Time Resolution Universe Survey, RIMAS and RATIR, and SKA South Africa/MeerKAT}),\ }\bibfield  {title} {\bibinfo {title} {{Multi-messenger Observations of a Binary Neutron Star Merger}},\ }\href {https://doi.org/10.3847/2041-8213/aa91c9} {\bibfield  {journal} {\bibinfo  {journal} {The Astrophysical Journal Letters}\ }\textbf {\bibinfo {volume} {848}},\ \bibinfo {pages} {L12} (\bibinfo {year} {2017}{\natexlab{b}})}\BibitemShut {NoStop}%
\bibitem [{\citenamefont {Abbott}\ \emph {et~al.}(2017{\natexlab{c}})\citenamefont {Abbott} \emph {et~al.}}]{GW170817_grb}%
  \BibitemOpen
  \bibfield  {author} {\bibinfo {author} {\bibfnamefont {B.~P.}\ \bibnamefont {Abbott}} \emph {et~al.} (\bibinfo {collaboration} {LIGO Scientific Collaboration and Virgo Collaboration, Fermi Gamma-ray Burst Monitor, INTEGRAL}),\ }\bibfield  {title} {\bibinfo {title} {{Gravitational Waves and Gamma-Rays from a Binary Neutron Star Merger: GW170817 and GRB 170817A}},\ }\href {https://doi.org/10.3847/2041-8213/aa920c} {\bibfield  {journal} {\bibinfo  {journal} {The Astrophysical Journal Letters}\ }\textbf {\bibinfo {volume} {848}},\ \bibinfo {pages} {L13} (\bibinfo {year} {2017}{\natexlab{c}})}\BibitemShut {NoStop}%
\bibitem [{\citenamefont {Abbott}\ \emph {et~al.}(2017{\natexlab{d}})\citenamefont {Abbott} \emph {et~al.}}]{Gw170817_cosmo}%
  \BibitemOpen
  \bibfield  {author} {\bibinfo {author} {\bibfnamefont {B.~P.}\ \bibnamefont {Abbott}} \emph {et~al.},\ }\bibfield  {title} {\bibinfo {title} {{A gravitational-wave standard siren measurement of the Hubble constant}},\ }\href {https://doi.org/10.1038/nature24471} {\bibfield  {journal} {\bibinfo  {journal} {Nature}\ }\textbf {\bibinfo {volume} {551}},\ \bibinfo {pages} {85–88} (\bibinfo {year} {2017}{\natexlab{d}})}\BibitemShut {NoStop}%
\bibitem [{\citenamefont {Abbott}\ \emph {et~al.}(2019{\natexlab{b}})\citenamefont {Abbott} \emph {et~al.}}]{Gw170817_GR}%
  \BibitemOpen
  \bibfield  {author} {\bibinfo {author} {\bibfnamefont {B.~P.}\ \bibnamefont {Abbott}} \emph {et~al.} (\bibinfo {collaboration} {LIGO Scientific Collaboration and Virgo Collaboration}),\ }\bibfield  {title} {\bibinfo {title} {{Tests of General Relativity with GW170817}},\ }\bibfield  {journal} {\bibinfo  {journal} {Physical Review Letters}\ }\textbf {\bibinfo {volume} {123}},\ \href {https://doi.org/10.1103/physrevlett.123.011102} {10.1103/physrevlett.123.011102} (\bibinfo {year} {2019}{\natexlab{b}})\BibitemShut {NoStop}%
\bibitem [{\citenamefont {Benhar}\ \emph {et~al.}(2024)\citenamefont {Benhar}, \citenamefont {Lovato}, \citenamefont {Maselli},\ and\ \citenamefont {Pannarale}}]{Benhar:2024qcw}%
  \BibitemOpen
  \bibinfo {editor} {\bibfnamefont {O.}~\bibnamefont {Benhar}}, \bibinfo {editor} {\bibfnamefont {A.}~\bibnamefont {Lovato}}, \bibinfo {editor} {\bibfnamefont {A.}~\bibnamefont {Maselli}},\ and\ \bibinfo {editor} {\bibfnamefont {F.}~\bibnamefont {Pannarale}},\ eds.,\ \href {https://doi.org/10.1201/9781003306580} {\emph {\bibinfo {title} {{Nuclear Theory in the Age of Multimessenger Astronomy}}}}\ (\bibinfo  {publisher} {CRC Press},\ \bibinfo {year} {2024})\BibitemShut {NoStop}%
\bibitem [{\citenamefont {Chatziioannou}\ \emph {et~al.}(2024)\citenamefont {Chatziioannou}, \citenamefont {Cromartie}, \citenamefont {Gandolfi}, \citenamefont {Tews}, \citenamefont {Radice}, \citenamefont {Steiner},\ and\ \citenamefont {Watts}}]{chatziioannou_2024}%
  \BibitemOpen
  \bibfield  {author} {\bibinfo {author} {\bibfnamefont {K.}~\bibnamefont {Chatziioannou}}, \bibinfo {author} {\bibfnamefont {H.~T.}\ \bibnamefont {Cromartie}}, \bibinfo {author} {\bibfnamefont {S.}~\bibnamefont {Gandolfi}}, \bibinfo {author} {\bibfnamefont {I.}~\bibnamefont {Tews}}, \bibinfo {author} {\bibfnamefont {D.}~\bibnamefont {Radice}}, \bibinfo {author} {\bibfnamefont {A.~W.}\ \bibnamefont {Steiner}},\ and\ \bibinfo {author} {\bibfnamefont {A.~L.}\ \bibnamefont {Watts}},\ }\href {https://arxiv.org/abs/2407.11153} {\bibinfo {title} {{Neutron stars and the dense matter equation of state: from microscopic theory to macroscopic observations}}} (\bibinfo {year} {2024}),\ \Eprint {https://arxiv.org/abs/2407.11153} {arXiv:2407.11153 [nucl-th]} \BibitemShut {NoStop}%
\bibitem [{\citenamefont {Chatziioannou}(2020)}]{Chatziioannou_2020}%
  \BibitemOpen
  \bibfield  {author} {\bibinfo {author} {\bibfnamefont {K.}~\bibnamefont {Chatziioannou}},\ }\bibfield  {title} {\bibinfo {title} {{Neutron-star tidal deformability and equation-of-state constraints}},\ }\bibfield  {journal} {\bibinfo  {journal} {General Relativity and Gravitation}\ }\textbf {\bibinfo {volume} {52}},\ \href {https://doi.org/10.1007/s10714-020-02754-3} {10.1007/s10714-020-02754-3} (\bibinfo {year} {2020})\BibitemShut {NoStop}%
\bibitem [{\citenamefont {Abbott}\ \emph {et~al.}(2017{\natexlab{e}})\citenamefont {Abbott} \emph {et~al.}}]{GW170817_postmerg}%
  \BibitemOpen
  \bibfield  {author} {\bibinfo {author} {\bibfnamefont {B.~P.}\ \bibnamefont {Abbott}} \emph {et~al.} (\bibinfo {collaboration} {LIGO Scientific Collaboration and Virgo Collaboration}),\ }\bibfield  {title} {\bibinfo {title} {Search for post-merger gravitational waves from the remnant of the binary neutron star merger gw170817},\ }\href {https://doi.org/10.3847/2041-8213/aa9a35} {\bibfield  {journal} {\bibinfo  {journal} {The Astrophysical Journal Letters}\ }\textbf {\bibinfo {volume} {851}},\ \bibinfo {pages} {L16} (\bibinfo {year} {2017}{\natexlab{e}})}\BibitemShut {NoStop}%
\bibitem [{\citenamefont {Flanagan}\ and\ \citenamefont {Hinderer}(2008)}]{Flanagan_2008}%
  \BibitemOpen
  \bibfield  {author} {\bibinfo {author} {\bibfnamefont {E.~E.}\ \bibnamefont {Flanagan}}\ and\ \bibinfo {author} {\bibfnamefont {T.}~\bibnamefont {Hinderer}},\ }\bibfield  {title} {\bibinfo {title} {Constraining neutron-star tidal love numbers with gravitational-wave detectors},\ }\href {https://doi.org/10.1103/PhysRevD.77.021502} {\bibfield  {journal} {\bibinfo  {journal} {Phys. Rev. D}\ }\textbf {\bibinfo {volume} {77}},\ \bibinfo {pages} {021502} (\bibinfo {year} {2008})}\BibitemShut {NoStop}%
\bibitem [{\citenamefont {Damour}\ and\ \citenamefont {Nagar}(2009)}]{Damour:2009vw}%
  \BibitemOpen
  \bibfield  {author} {\bibinfo {author} {\bibfnamefont {T.}~\bibnamefont {Damour}}\ and\ \bibinfo {author} {\bibfnamefont {A.}~\bibnamefont {Nagar}},\ }\bibfield  {title} {\bibinfo {title} {{Relativistic tidal properties of neutron stars}},\ }\href {https://doi.org/10.1103/PhysRevD.80.084035} {\bibfield  {journal} {\bibinfo  {journal} {Phys. Rev. D}\ }\textbf {\bibinfo {volume} {80}},\ \bibinfo {pages} {084035} (\bibinfo {year} {2009})},\ \Eprint {https://arxiv.org/abs/0906.0096} {arXiv:0906.0096 [gr-qc]} \BibitemShut {NoStop}%
\bibitem [{\citenamefont {Abbott}\ \emph {et~al.}(2020{\natexlab{b}})\citenamefont {Abbott} \emph {et~al.}}]{LVKguide}%
  \BibitemOpen
  \bibfield  {author} {\bibinfo {author} {\bibfnamefont {B.}~\bibnamefont {Abbott}} \emph {et~al.} (\bibinfo {collaboration} {LIGO Scientific Collaboration and Virgo Collaboration}),\ }\bibfield  {title} {\bibinfo {title} {{A guide to LIGO–Virgo detector noise and extraction of transient gravitational-wave signals}},\ }\href {https://doi.org/10.1088/1361-6382/ab685e} {\bibfield  {journal} {\bibinfo  {journal} {Classical and Quantum Gravity}\ }\textbf {\bibinfo {volume} {37}},\ \bibinfo {pages} {055002} (\bibinfo {year} {2020}{\natexlab{b}})}\BibitemShut {NoStop}%
\bibitem [{\citenamefont {Usman}\ \emph {et~al.}(2016)\citenamefont {Usman}, \citenamefont {Nitz}, \citenamefont {Harry}, \citenamefont {Biwer}, \citenamefont {Brown}, \citenamefont {Cabero}, \citenamefont {Capano}, \citenamefont {Canton}, \citenamefont {Dent}, \citenamefont {Fairhurst}, \citenamefont {Kehl}, \citenamefont {Keppel}, \citenamefont {Krishnan}, \citenamefont {Lenon}, \citenamefont {Lundgren}, \citenamefont {Nielsen}, \citenamefont {Pekowsky}, \citenamefont {Pfeiffer}, \citenamefont {Saulson}, \citenamefont {West},\ and\ \citenamefont {Willis}}]{pycbc_guide}%
  \BibitemOpen
  \bibfield  {author} {\bibinfo {author} {\bibfnamefont {S.~A.}\ \bibnamefont {Usman}}, \bibinfo {author} {\bibfnamefont {A.~H.}\ \bibnamefont {Nitz}}, \bibinfo {author} {\bibfnamefont {I.~W.}\ \bibnamefont {Harry}}, \bibinfo {author} {\bibfnamefont {C.~M.}\ \bibnamefont {Biwer}}, \bibinfo {author} {\bibfnamefont {D.~A.}\ \bibnamefont {Brown}}, \bibinfo {author} {\bibfnamefont {M.}~\bibnamefont {Cabero}}, \bibinfo {author} {\bibfnamefont {C.~D.}\ \bibnamefont {Capano}}, \bibinfo {author} {\bibfnamefont {T.~D.}\ \bibnamefont {Canton}}, \bibinfo {author} {\bibfnamefont {T.}~\bibnamefont {Dent}}, \bibinfo {author} {\bibfnamefont {S.}~\bibnamefont {Fairhurst}}, \bibinfo {author} {\bibfnamefont {M.~S.}\ \bibnamefont {Kehl}}, \bibinfo {author} {\bibfnamefont {D.}~\bibnamefont {Keppel}}, \bibinfo {author} {\bibfnamefont {B.}~\bibnamefont {Krishnan}}, \bibinfo {author} {\bibfnamefont {A.}~\bibnamefont {Lenon}}, \bibinfo {author} {\bibfnamefont {A.}~\bibnamefont {Lundgren}}, \bibinfo {author} {\bibfnamefont {A.~B.}\
  \bibnamefont {Nielsen}}, \bibinfo {author} {\bibfnamefont {L.~P.}\ \bibnamefont {Pekowsky}}, \bibinfo {author} {\bibfnamefont {H.~P.}\ \bibnamefont {Pfeiffer}}, \bibinfo {author} {\bibfnamefont {P.~R.}\ \bibnamefont {Saulson}}, \bibinfo {author} {\bibfnamefont {M.}~\bibnamefont {West}},\ and\ \bibinfo {author} {\bibfnamefont {J.~L.}\ \bibnamefont {Willis}},\ }\bibfield  {title} {\bibinfo {title} {{The PyCBC search for gravitational waves from compact binary coalescence}},\ }\href {https://doi.org/10.1088/0264-9381/33/21/215004} {\bibfield  {journal} {\bibinfo  {journal} {Classical and Quantum Gravity}\ }\textbf {\bibinfo {volume} {33}},\ \bibinfo {pages} {215004} (\bibinfo {year} {2016})}\BibitemShut {NoStop}%
\bibitem [{\citenamefont {Owen}(1996)}]{Owen_1996}%
  \BibitemOpen
  \bibfield  {author} {\bibinfo {author} {\bibfnamefont {B.~J.}\ \bibnamefont {Owen}},\ }\bibfield  {title} {\bibinfo {title} {{Search templates for gravitational waves from inspiraling binaries: Choice of template spacing}},\ }\href {https://doi.org/10.1103/PhysRevD.53.6749} {\bibfield  {journal} {\bibinfo  {journal} {Phys. Rev. D}\ }\textbf {\bibinfo {volume} {53}},\ \bibinfo {pages} {6749} (\bibinfo {year} {1996})}\BibitemShut {NoStop}%
\bibitem [{\citenamefont {Hinderer}\ \emph {et~al.}(2010)\citenamefont {Hinderer}, \citenamefont {Lackey}, \citenamefont {Lang},\ and\ \citenamefont {Read}}]{Hinderer_2009}%
  \BibitemOpen
  \bibfield  {author} {\bibinfo {author} {\bibfnamefont {T.}~\bibnamefont {Hinderer}}, \bibinfo {author} {\bibfnamefont {B.~D.}\ \bibnamefont {Lackey}}, \bibinfo {author} {\bibfnamefont {R.~N.}\ \bibnamefont {Lang}},\ and\ \bibinfo {author} {\bibfnamefont {J.~S.}\ \bibnamefont {Read}},\ }\bibfield  {title} {\bibinfo {title} {{Tidal deformability of neutron stars with realistic equations of state and their gravitational wave signatures in binary inspiral}},\ }\href {https://doi.org/10.1103/PhysRevD.81.123016} {\bibfield  {journal} {\bibinfo  {journal} {Phys. Rev. D}\ }\textbf {\bibinfo {volume} {81}},\ \bibinfo {pages} {123016} (\bibinfo {year} {2010})}\BibitemShut {NoStop}%
\bibitem [{\citenamefont {Cullen}\ \emph {et~al.}(2017)\citenamefont {Cullen}, \citenamefont {Harry}, \citenamefont {Read},\ and\ \citenamefont {Flynn}}]{Cullen_2017}%
  \BibitemOpen
  \bibfield  {author} {\bibinfo {author} {\bibfnamefont {T.}~\bibnamefont {Cullen}}, \bibinfo {author} {\bibfnamefont {I.}~\bibnamefont {Harry}}, \bibinfo {author} {\bibfnamefont {J.}~\bibnamefont {Read}},\ and\ \bibinfo {author} {\bibfnamefont {E.}~\bibnamefont {Flynn}},\ }\bibfield  {title} {\bibinfo {title} {{Matter effects on LIGO/Virgo searches for gravitational waves from merging neutron stars}},\ }\href {https://doi.org/10.1088/1361-6382/aa9424} {\bibfield  {journal} {\bibinfo  {journal} {Classical and Quantum Gravity}\ }\textbf {\bibinfo {volume} {34}},\ \bibinfo {pages} {245003} (\bibinfo {year} {2017})}\BibitemShut {NoStop}%
\bibitem [{\citenamefont {Harry}\ and\ \citenamefont {Hinderer}(2018)}]{Harry_2018}%
  \BibitemOpen
  \bibfield  {author} {\bibinfo {author} {\bibfnamefont {I.}~\bibnamefont {Harry}}\ and\ \bibinfo {author} {\bibfnamefont {T.}~\bibnamefont {Hinderer}},\ }\bibfield  {title} {\bibinfo {title} {{Observing and measuring the neutron-star equation-of-state in spinning binary neutron star systems}},\ }\href {https://doi.org/10.1088/1361-6382/aac7e3} {\bibfield  {journal} {\bibinfo  {journal} {Classical and Quantum Gravity}\ }\textbf {\bibinfo {volume} {35}},\ \bibinfo {pages} {145010} (\bibinfo {year} {2018})}\BibitemShut {NoStop}%
\bibitem [{\citenamefont {Allen}(2021)}]{Allen_2021}%
  \BibitemOpen
  \bibfield  {author} {\bibinfo {author} {\bibfnamefont {B.}~\bibnamefont {Allen}},\ }\bibfield  {title} {\bibinfo {title} {{Optimal template banks}},\ }\href {https://doi.org/10.1103/PhysRevD.104.042005} {\bibfield  {journal} {\bibinfo  {journal} {Phys. Rev. D}\ }\textbf {\bibinfo {volume} {104}},\ \bibinfo {pages} {042005} (\bibinfo {year} {2021})}\BibitemShut {NoStop}%
\bibitem [{\citenamefont {Radice}\ \emph {et~al.}(2018)\citenamefont {Radice}, \citenamefont {Perego}, \citenamefont {Zappa},\ and\ \citenamefont {Bernuzzi}}]{Radice_2018}%
  \BibitemOpen
  \bibfield  {author} {\bibinfo {author} {\bibfnamefont {D.}~\bibnamefont {Radice}}, \bibinfo {author} {\bibfnamefont {A.}~\bibnamefont {Perego}}, \bibinfo {author} {\bibfnamefont {F.}~\bibnamefont {Zappa}},\ and\ \bibinfo {author} {\bibfnamefont {S.}~\bibnamefont {Bernuzzi}},\ }\bibfield  {title} {\bibinfo {title} {{GW170817: Joint Constraint on the Neutron Star Equation of State from Multimessenger Observations}},\ }\href {https://doi.org/10.3847/2041-8213/aaa402} {\bibfield  {journal} {\bibinfo  {journal} {The Astrophysical Journal Letters}\ }\textbf {\bibinfo {volume} {852}},\ \bibinfo {pages} {L29} (\bibinfo {year} {2018})}\BibitemShut {NoStop}%
\bibitem [{\citenamefont {Radice}\ and\ \citenamefont {Dai}(2019)}]{Radice_2019}%
  \BibitemOpen
  \bibfield  {author} {\bibinfo {author} {\bibfnamefont {D.}~\bibnamefont {Radice}}\ and\ \bibinfo {author} {\bibfnamefont {L.}~\bibnamefont {Dai}},\ }\bibfield  {title} {\bibinfo {title} {{Multimessenger parameter estimation of GW170817}},\ }\bibfield  {journal} {\bibinfo  {journal} {The European Physical Journal A}\ }\textbf {\bibinfo {volume} {55}},\ \href {https://doi.org/10.1140/epja/i2019-12716-4} {10.1140/epja/i2019-12716-4} (\bibinfo {year} {2019})\BibitemShut {NoStop}%
\bibitem [{\citenamefont {Harry}\ \emph {et~al.}(2009)\citenamefont {Harry}, \citenamefont {Allen},\ and\ \citenamefont {Sathyaprakash}}]{Harry_2009}%
  \BibitemOpen
  \bibfield  {author} {\bibinfo {author} {\bibfnamefont {I.~W.}\ \bibnamefont {Harry}}, \bibinfo {author} {\bibfnamefont {B.}~\bibnamefont {Allen}},\ and\ \bibinfo {author} {\bibfnamefont {B.~S.}\ \bibnamefont {Sathyaprakash}},\ }\bibfield  {title} {\bibinfo {title} {{Stochastic template placement algorithm for gravitational wave data analysis}},\ }\href {https://doi.org/10.1103/PhysRevD.80.104014} {\bibfield  {journal} {\bibinfo  {journal} {Phys. Rev. D}\ }\textbf {\bibinfo {volume} {80}},\ \bibinfo {pages} {104014} (\bibinfo {year} {2009})}\BibitemShut {NoStop}%
\bibitem [{\citenamefont {Kacanja}\ \emph {et~al.}(2024)\citenamefont {Kacanja}, \citenamefont {Nitz}, \citenamefont {Wu}, \citenamefont {Cusinato}, \citenamefont {Dhurkunde}, \citenamefont {Harry}, \citenamefont {Canton},\ and\ \citenamefont {Pannarale}}]{kacanja_2024}%
  \BibitemOpen
  \bibfield  {author} {\bibinfo {author} {\bibfnamefont {K.}~\bibnamefont {Kacanja}}, \bibinfo {author} {\bibfnamefont {A.~H.}\ \bibnamefont {Nitz}}, \bibinfo {author} {\bibfnamefont {S.}~\bibnamefont {Wu}}, \bibinfo {author} {\bibfnamefont {M.}~\bibnamefont {Cusinato}}, \bibinfo {author} {\bibfnamefont {R.}~\bibnamefont {Dhurkunde}}, \bibinfo {author} {\bibfnamefont {I.}~\bibnamefont {Harry}}, \bibinfo {author} {\bibfnamefont {T.~D.}\ \bibnamefont {Canton}},\ and\ \bibinfo {author} {\bibfnamefont {F.}~\bibnamefont {Pannarale}},\ }\href {https://arxiv.org/abs/2407.03406} {\bibinfo {title} {{Efficient Stochastic Template Bank using Inner Product Inequalities}}} (\bibinfo {year} {2024}),\ \Eprint {https://arxiv.org/abs/2407.03406} {arXiv:2407.03406 [astro-ph.HE]} \BibitemShut {NoStop}%
\bibitem [{\citenamefont {Harry}\ and\ \citenamefont {Lundgren}(2021)}]{Harry_Lundgren}%
  \BibitemOpen
  \bibfield  {author} {\bibinfo {author} {\bibfnamefont {I.}~\bibnamefont {Harry}}\ and\ \bibinfo {author} {\bibfnamefont {A.}~\bibnamefont {Lundgren}},\ }\bibfield  {title} {\bibinfo {title} {{Failure of the Fisher matrix when including tidal terms: Considering construction of template banks of tidally deformed binary neutron stars}},\ }\href {https://doi.org/10.1103/PhysRevD.104.043008} {\bibfield  {journal} {\bibinfo  {journal} {Phys. Rev. D}\ }\textbf {\bibinfo {volume} {104}},\ \bibinfo {pages} {043008} (\bibinfo {year} {2021})}\BibitemShut {NoStop}%
\bibitem [{\citenamefont {Nitz}\ \emph {et~al.}(2024)\citenamefont {Nitz}, \citenamefont {Harry}, \citenamefont {Brown}, \citenamefont {Biwer}, \citenamefont {Willis}, \citenamefont {Canton}, \citenamefont {Capano}, \citenamefont {Dent}, \citenamefont {Pekowsky}, \citenamefont {Davies}, \citenamefont {De}, \citenamefont {Cabero}, \citenamefont {Wu}, \citenamefont {Williamson}, \citenamefont {Machenschalk}, \citenamefont {Macleod}, \citenamefont {Pannarale}, \citenamefont {Kumar}, \citenamefont {Reyes}, \citenamefont {dfinstad}, \citenamefont {Kumar}, \citenamefont {Tápai}, \citenamefont {Singer}, \citenamefont {Kumar}, \citenamefont {veronica villa}, \citenamefont {maxtrevor}, \citenamefont {Gadre}, \citenamefont {Khan}, \citenamefont {Fairhurst},\ and\ \citenamefont {Tolley}}]{PyCBC}%
  \BibitemOpen
  \bibfield  {author} {\bibinfo {author} {\bibfnamefont {A.}~\bibnamefont {Nitz}}, \bibinfo {author} {\bibfnamefont {I.}~\bibnamefont {Harry}}, \bibinfo {author} {\bibfnamefont {D.}~\bibnamefont {Brown}}, \bibinfo {author} {\bibfnamefont {C.~M.}\ \bibnamefont {Biwer}}, \bibinfo {author} {\bibfnamefont {J.}~\bibnamefont {Willis}}, \bibinfo {author} {\bibfnamefont {T.~D.}\ \bibnamefont {Canton}}, \bibinfo {author} {\bibfnamefont {C.}~\bibnamefont {Capano}}, \bibinfo {author} {\bibfnamefont {T.}~\bibnamefont {Dent}}, \bibinfo {author} {\bibfnamefont {L.}~\bibnamefont {Pekowsky}}, \bibinfo {author} {\bibfnamefont {G.~S.~C.}\ \bibnamefont {Davies}}, \bibinfo {author} {\bibfnamefont {S.}~\bibnamefont {De}}, \bibinfo {author} {\bibfnamefont {M.}~\bibnamefont {Cabero}}, \bibinfo {author} {\bibfnamefont {S.}~\bibnamefont {Wu}}, \bibinfo {author} {\bibfnamefont {A.~R.}\ \bibnamefont {Williamson}}, \bibinfo {author} {\bibfnamefont {B.}~\bibnamefont {Machenschalk}}, \bibinfo {author} {\bibfnamefont {D.}~\bibnamefont
  {Macleod}}, \bibinfo {author} {\bibfnamefont {F.}~\bibnamefont {Pannarale}}, \bibinfo {author} {\bibfnamefont {P.}~\bibnamefont {Kumar}}, \bibinfo {author} {\bibfnamefont {S.}~\bibnamefont {Reyes}}, \bibinfo {author} {\bibnamefont {dfinstad}}, \bibinfo {author} {\bibfnamefont {S.}~\bibnamefont {Kumar}}, \bibinfo {author} {\bibfnamefont {M.}~\bibnamefont {Tápai}}, \bibinfo {author} {\bibfnamefont {L.}~\bibnamefont {Singer}}, \bibinfo {author} {\bibfnamefont {P.}~\bibnamefont {Kumar}}, \bibinfo {author} {\bibnamefont {veronica villa}}, \bibinfo {author} {\bibnamefont {maxtrevor}}, \bibinfo {author} {\bibfnamefont {B.~U.~V.}\ \bibnamefont {Gadre}}, \bibinfo {author} {\bibfnamefont {S.}~\bibnamefont {Khan}}, \bibinfo {author} {\bibfnamefont {S.}~\bibnamefont {Fairhurst}},\ and\ \bibinfo {author} {\bibfnamefont {A.}~\bibnamefont {Tolley}},\ }\href {https://doi.org/10.5281/zenodo.10473621} {\bibinfo {title} {{gwastro/pycbc: v2.3.3 release of PyCBC}}} (\bibinfo {year} {2024})\BibitemShut {NoStop}%
\bibitem [{Note2()}]{Note2}%
  \BibitemOpen
  \bibinfo {note} {\protect \url {https://dcc.ligo.org/LIGO-T2200043/public}}\BibitemShut {NoStop}%
\bibitem [{\citenamefont {Roy}\ \emph {et~al.}(2017)\citenamefont {Roy}, \citenamefont {Sengupta},\ and\ \citenamefont {Thakor}}]{Roy:2017qgg}%
  \BibitemOpen
  \bibfield  {author} {\bibinfo {author} {\bibfnamefont {S.}~\bibnamefont {Roy}}, \bibinfo {author} {\bibfnamefont {A.~S.}\ \bibnamefont {Sengupta}},\ and\ \bibinfo {author} {\bibfnamefont {N.}~\bibnamefont {Thakor}},\ }\bibfield  {title} {\bibinfo {title} {{Hybrid geometric-random template-placement algorithm for gravitational wave searches from compact binary coalescences}},\ }\href {https://doi.org/10.1103/PhysRevD.95.104045} {\bibfield  {journal} {\bibinfo  {journal} {Phys. Rev. D}\ }\textbf {\bibinfo {volume} {95}},\ \bibinfo {pages} {104045} (\bibinfo {year} {2017})},\ \Eprint {https://arxiv.org/abs/1702.06771} {arXiv:1702.06771 [gr-qc]} \BibitemShut {NoStop}%
\bibitem [{\citenamefont {Roy}\ \emph {et~al.}(2019)\citenamefont {Roy}, \citenamefont {Sengupta},\ and\ \citenamefont {Ajith}}]{Roy:2017oul}%
  \BibitemOpen
  \bibfield  {author} {\bibinfo {author} {\bibfnamefont {S.}~\bibnamefont {Roy}}, \bibinfo {author} {\bibfnamefont {A.~S.}\ \bibnamefont {Sengupta}},\ and\ \bibinfo {author} {\bibfnamefont {P.}~\bibnamefont {Ajith}},\ }\bibfield  {title} {\bibinfo {title} {{Effectual template banks for upcoming compact binary searches in Advanced-LIGO and Virgo data}},\ }\href {https://doi.org/10.1103/PhysRevD.99.024048} {\bibfield  {journal} {\bibinfo  {journal} {Phys. Rev. D}\ }\textbf {\bibinfo {volume} {99}},\ \bibinfo {pages} {024048} (\bibinfo {year} {2019})},\ \Eprint {https://arxiv.org/abs/1711.08743} {arXiv:1711.08743 [gr-qc]} \BibitemShut {NoStop}%
\bibitem [{\citenamefont {Buonanno}\ and\ \citenamefont {Damour}(1999)}]{Buonanno_1999}%
  \BibitemOpen
  \bibfield  {author} {\bibinfo {author} {\bibfnamefont {A.}~\bibnamefont {Buonanno}}\ and\ \bibinfo {author} {\bibfnamefont {T.}~\bibnamefont {Damour}},\ }\bibfield  {title} {\bibinfo {title} {{Effective one-body approach to general relativistic two-body dynamics}},\ }\href {https://doi.org/10.1103/PhysRevD.59.084006} {\bibfield  {journal} {\bibinfo  {journal} {Phys. Rev. D}\ }\textbf {\bibinfo {volume} {59}},\ \bibinfo {pages} {084006} (\bibinfo {year} {1999})}\BibitemShut {NoStop}%
\bibitem [{\citenamefont {Buonanno}\ \emph {et~al.}(2007)\citenamefont {Buonanno}, \citenamefont {Pan}, \citenamefont {Baker}, \citenamefont {Centrella}, \citenamefont {Kelly}, \citenamefont {McWilliams},\ and\ \citenamefont {van Meter}}]{Buonanno_2007}%
  \BibitemOpen
  \bibfield  {author} {\bibinfo {author} {\bibfnamefont {A.}~\bibnamefont {Buonanno}}, \bibinfo {author} {\bibfnamefont {Y.}~\bibnamefont {Pan}}, \bibinfo {author} {\bibfnamefont {J.~G.}\ \bibnamefont {Baker}}, \bibinfo {author} {\bibfnamefont {J.}~\bibnamefont {Centrella}}, \bibinfo {author} {\bibfnamefont {B.~J.}\ \bibnamefont {Kelly}}, \bibinfo {author} {\bibfnamefont {S.~T.}\ \bibnamefont {McWilliams}},\ and\ \bibinfo {author} {\bibfnamefont {J.~R.}\ \bibnamefont {van Meter}},\ }\bibfield  {title} {\bibinfo {title} {{Approaching faithful templates for nonspinning binary black holes using the effective-one-body approach}},\ }\href {https://doi.org/10.1103/PhysRevD.76.104049} {\bibfield  {journal} {\bibinfo  {journal} {Phys. Rev. D}\ }\textbf {\bibinfo {volume} {76}},\ \bibinfo {pages} {104049} (\bibinfo {year} {2007})}\BibitemShut {NoStop}%
\bibitem [{\citenamefont {Boh\'e}\ \emph {et~al.}(2017)\citenamefont {Boh\'e}, \citenamefont {Shao}, \citenamefont {Taracchini}, \citenamefont {Buonanno}, \citenamefont {Babak}, \citenamefont {Harry}, \citenamefont {Hinder}, \citenamefont {Ossokine}, \citenamefont {P\"urrer}, \citenamefont {Raymond}, \citenamefont {Chu}, \citenamefont {Fong}, \citenamefont {Kumar}, \citenamefont {Pfeiffer}, \citenamefont {Boyle}, \citenamefont {Hemberger}, \citenamefont {Kidder}, \citenamefont {Lovelace}, \citenamefont {Scheel},\ and\ \citenamefont {Szil\'agyi}}]{Bohe_2017}%
  \BibitemOpen
  \bibfield  {author} {\bibinfo {author} {\bibfnamefont {A.}~\bibnamefont {Boh\'e}}, \bibinfo {author} {\bibfnamefont {L.}~\bibnamefont {Shao}}, \bibinfo {author} {\bibfnamefont {A.}~\bibnamefont {Taracchini}}, \bibinfo {author} {\bibfnamefont {A.}~\bibnamefont {Buonanno}}, \bibinfo {author} {\bibfnamefont {S.}~\bibnamefont {Babak}}, \bibinfo {author} {\bibfnamefont {I.~W.}\ \bibnamefont {Harry}}, \bibinfo {author} {\bibfnamefont {I.}~\bibnamefont {Hinder}}, \bibinfo {author} {\bibfnamefont {S.}~\bibnamefont {Ossokine}}, \bibinfo {author} {\bibfnamefont {M.}~\bibnamefont {P\"urrer}}, \bibinfo {author} {\bibfnamefont {V.}~\bibnamefont {Raymond}}, \bibinfo {author} {\bibfnamefont {T.}~\bibnamefont {Chu}}, \bibinfo {author} {\bibfnamefont {H.}~\bibnamefont {Fong}}, \bibinfo {author} {\bibfnamefont {P.}~\bibnamefont {Kumar}}, \bibinfo {author} {\bibfnamefont {H.~P.}\ \bibnamefont {Pfeiffer}}, \bibinfo {author} {\bibfnamefont {M.}~\bibnamefont {Boyle}}, \bibinfo {author} {\bibfnamefont {D.~A.}\ \bibnamefont
  {Hemberger}}, \bibinfo {author} {\bibfnamefont {L.~E.}\ \bibnamefont {Kidder}}, \bibinfo {author} {\bibfnamefont {G.}~\bibnamefont {Lovelace}}, \bibinfo {author} {\bibfnamefont {M.~A.}\ \bibnamefont {Scheel}},\ and\ \bibinfo {author} {\bibfnamefont {B.}~\bibnamefont {Szil\'agyi}},\ }\bibfield  {title} {\bibinfo {title} {{Improved effective-one-body model of spinning, nonprecessing binary black holes for the era of gravitational-wave astrophysics with advanced detectors}},\ }\href {https://doi.org/10.1103/PhysRevD.95.044028} {\bibfield  {journal} {\bibinfo  {journal} {Phys. Rev. D}\ }\textbf {\bibinfo {volume} {95}},\ \bibinfo {pages} {044028} (\bibinfo {year} {2017})}\BibitemShut {NoStop}%
\bibitem [{\citenamefont {Pompili}\ \emph {et~al.}(2023)\citenamefont {Pompili}, \citenamefont {Buonanno}, \citenamefont {Estell\'es}, \citenamefont {Khalil}, \citenamefont {van~de Meent}, \citenamefont {Mihaylov}, \citenamefont {Ossokine}, \citenamefont {P\"urrer}, \citenamefont {Ramos-Buades}, \citenamefont {Mehta}, \citenamefont {Cotesta}, \citenamefont {Marsat}, \citenamefont {Boyle}, \citenamefont {Kidder}, \citenamefont {Pfeiffer}, \citenamefont {Scheel}, \citenamefont {R\"uter}, \citenamefont {Vu}, \citenamefont {Dudi}, \citenamefont {Ma}, \citenamefont {Mitman}, \citenamefont {Melchor}, \citenamefont {Thomas},\ and\ \citenamefont {Sanchez}}]{Pompili_2023}%
  \BibitemOpen
  \bibfield  {author} {\bibinfo {author} {\bibfnamefont {L.}~\bibnamefont {Pompili}}, \bibinfo {author} {\bibfnamefont {A.}~\bibnamefont {Buonanno}}, \bibinfo {author} {\bibfnamefont {H.}~\bibnamefont {Estell\'es}}, \bibinfo {author} {\bibfnamefont {M.}~\bibnamefont {Khalil}}, \bibinfo {author} {\bibfnamefont {M.}~\bibnamefont {van~de Meent}}, \bibinfo {author} {\bibfnamefont {D.~P.}\ \bibnamefont {Mihaylov}}, \bibinfo {author} {\bibfnamefont {S.}~\bibnamefont {Ossokine}}, \bibinfo {author} {\bibfnamefont {M.}~\bibnamefont {P\"urrer}}, \bibinfo {author} {\bibfnamefont {A.}~\bibnamefont {Ramos-Buades}}, \bibinfo {author} {\bibfnamefont {A.~K.}\ \bibnamefont {Mehta}}, \bibinfo {author} {\bibfnamefont {R.}~\bibnamefont {Cotesta}}, \bibinfo {author} {\bibfnamefont {S.}~\bibnamefont {Marsat}}, \bibinfo {author} {\bibfnamefont {M.}~\bibnamefont {Boyle}}, \bibinfo {author} {\bibfnamefont {L.~E.}\ \bibnamefont {Kidder}}, \bibinfo {author} {\bibfnamefont {H.~P.}\ \bibnamefont {Pfeiffer}}, \bibinfo {author}
  {\bibfnamefont {M.~A.}\ \bibnamefont {Scheel}}, \bibinfo {author} {\bibfnamefont {H.~R.}\ \bibnamefont {R\"uter}}, \bibinfo {author} {\bibfnamefont {N.}~\bibnamefont {Vu}}, \bibinfo {author} {\bibfnamefont {R.}~\bibnamefont {Dudi}}, \bibinfo {author} {\bibfnamefont {S.}~\bibnamefont {Ma}}, \bibinfo {author} {\bibfnamefont {K.}~\bibnamefont {Mitman}}, \bibinfo {author} {\bibfnamefont {D.}~\bibnamefont {Melchor}}, \bibinfo {author} {\bibfnamefont {S.}~\bibnamefont {Thomas}},\ and\ \bibinfo {author} {\bibfnamefont {J.}~\bibnamefont {Sanchez}},\ }\bibfield  {title} {\bibinfo {title} {{Laying the foundation of the effective-one-body waveform models SEOBNRv5: Improved accuracy and efficiency for spinning nonprecessing binary black holes}},\ }\href {https://doi.org/10.1103/PhysRevD.108.124035} {\bibfield  {journal} {\bibinfo  {journal} {Phys. Rev. D}\ }\textbf {\bibinfo {volume} {108}},\ \bibinfo {pages} {124035} (\bibinfo {year} {2023})}\BibitemShut {NoStop}%
\bibitem [{\citenamefont {Chabanat}\ \emph {et~al.}(1998)\citenamefont {Chabanat}, \citenamefont {Bonche}, \citenamefont {Haensel}, \citenamefont {Meyer},\ and\ \citenamefont {Schaeffer}}]{Chabanat_1998}%
  \BibitemOpen
  \bibfield  {author} {\bibinfo {author} {\bibfnamefont {E.}~\bibnamefont {Chabanat}}, \bibinfo {author} {\bibfnamefont {P.}~\bibnamefont {Bonche}}, \bibinfo {author} {\bibfnamefont {P.}~\bibnamefont {Haensel}}, \bibinfo {author} {\bibfnamefont {J.}~\bibnamefont {Meyer}},\ and\ \bibinfo {author} {\bibfnamefont {R.}~\bibnamefont {Schaeffer}},\ }\bibfield  {title} {\bibinfo {title} {{A Skyrme parametrization from subnuclear to neutron star densities Part II. Nuclei far from stabilities}},\ }\href {https://doi.org/https://doi.org/10.1016/S0375-9474(98)00180-8} {\bibfield  {journal} {\bibinfo  {journal} {Nuclear Physics A}\ }\textbf {\bibinfo {volume} {635}},\ \bibinfo {pages} {231} (\bibinfo {year} {1998})}\BibitemShut {NoStop}%
\bibitem [{\citenamefont {Douchin}\ and\ \citenamefont {Haensel}(2001)}]{Douchin_2001}%
  \BibitemOpen
  \bibfield  {author} {\bibinfo {author} {\bibfnamefont {F.}~\bibnamefont {Douchin}}\ and\ \bibinfo {author} {\bibfnamefont {P.}~\bibnamefont {Haensel}},\ }\bibfield  {title} {\bibinfo {title} {{A unified equation of state of dense matter and neutron star structure}},\ }\href {https://doi.org/10.1051/0004-6361:20011402} {\bibfield  {journal} {\bibinfo  {journal} {Astronomy \& Astrophysics}\ }\textbf {\bibinfo {volume} {380}},\ \bibinfo {pages} {151–167} (\bibinfo {year} {2001})}\BibitemShut {NoStop}%
\bibitem [{\citenamefont {Müther}\ \emph {et~al.}(1987)\citenamefont {Müther}, \citenamefont {Prakash},\ and\ \citenamefont {Ainsworth}}]{Muther_1987}%
  \BibitemOpen
  \bibfield  {author} {\bibinfo {author} {\bibfnamefont {H.}~\bibnamefont {Müther}}, \bibinfo {author} {\bibfnamefont {M.}~\bibnamefont {Prakash}},\ and\ \bibinfo {author} {\bibfnamefont {T.}~\bibnamefont {Ainsworth}},\ }\bibfield  {title} {\bibinfo {title} {{The nuclear symmetry energy in relativistic Brueckner-Hartree-Fock calculations}},\ }\href {https://doi.org/https://doi.org/10.1016/0370-2693(87)91611-X} {\bibfield  {journal} {\bibinfo  {journal} {Physics Letters B}\ }\textbf {\bibinfo {volume} {199}},\ \bibinfo {pages} {469} (\bibinfo {year} {1987})}\BibitemShut {NoStop}%
\bibitem [{\citenamefont {Friedrich}\ and\ \citenamefont {Reinhard}(1986)}]{friedrich_1986}%
  \BibitemOpen
  \bibfield  {author} {\bibinfo {author} {\bibfnamefont {J.}~\bibnamefont {Friedrich}}\ and\ \bibinfo {author} {\bibfnamefont {P.-G.}\ \bibnamefont {Reinhard}},\ }\bibfield  {title} {\bibinfo {title} {{Skyrme-force parametrization: Least-squares fit to nuclear ground-state properties}},\ }\href {https://doi.org/10.1103/PhysRevC.33.335} {\bibfield  {journal} {\bibinfo  {journal} {Phys. Rev. C}\ }\textbf {\bibinfo {volume} {33}},\ \bibinfo {pages} {335} (\bibinfo {year} {1986})}\BibitemShut {NoStop}%
\bibitem [{\citenamefont {Danielewicz}\ and\ \citenamefont {Lee}(2009)}]{DANIELEWICZ200936}%
  \BibitemOpen
  \bibfield  {author} {\bibinfo {author} {\bibfnamefont {P.}~\bibnamefont {Danielewicz}}\ and\ \bibinfo {author} {\bibfnamefont {J.}~\bibnamefont {Lee}},\ }\bibfield  {title} {\bibinfo {title} {{Symmetry energy I: Semi-infinite matter}},\ }\href {https://doi.org/https://doi.org/10.1016/j.nuclphysa.2008.11.007} {\bibfield  {journal} {\bibinfo  {journal} {Nuclear Physics A}\ }\textbf {\bibinfo {volume} {818}},\ \bibinfo {pages} {36} (\bibinfo {year} {2009})}\BibitemShut {NoStop}%
\bibitem [{\citenamefont {Gulminelli}\ and\ \citenamefont {Raduta}(2015)}]{Gulminelli_2015}%
  \BibitemOpen
  \bibfield  {author} {\bibinfo {author} {\bibfnamefont {F.}~\bibnamefont {Gulminelli}}\ and\ \bibinfo {author} {\bibfnamefont {A.~R.}\ \bibnamefont {Raduta}},\ }\bibfield  {title} {\bibinfo {title} {{Unified treatment of subsaturation stellar matter at zero and finite temperature}},\ }\href {https://doi.org/10.1103/PhysRevC.92.055803} {\bibfield  {journal} {\bibinfo  {journal} {Phys. Rev. C}\ }\textbf {\bibinfo {volume} {92}},\ \bibinfo {pages} {055803} (\bibinfo {year} {2015})}\BibitemShut {NoStop}%
\bibitem [{\citenamefont {Typel}\ \emph {et~al.}(2010)\citenamefont {Typel}, \citenamefont {Röpke}, \citenamefont {Klähn}, \citenamefont {Blaschke},\ and\ \citenamefont {Wolter}}]{Typel_2010}%
  \BibitemOpen
  \bibfield  {author} {\bibinfo {author} {\bibfnamefont {S.}~\bibnamefont {Typel}}, \bibinfo {author} {\bibfnamefont {G.}~\bibnamefont {Röpke}}, \bibinfo {author} {\bibfnamefont {T.}~\bibnamefont {Klähn}}, \bibinfo {author} {\bibfnamefont {D.}~\bibnamefont {Blaschke}},\ and\ \bibinfo {author} {\bibfnamefont {H.~H.}\ \bibnamefont {Wolter}},\ }\bibfield  {title} {\bibinfo {title} {{Composition and thermodynamics of nuclear matter with light clusters}},\ }\bibfield  {journal} {\bibinfo  {journal} {Physical Review C}\ }\textbf {\bibinfo {volume} {81}},\ \href {https://doi.org/10.1103/physrevc.81.015803} {10.1103/physrevc.81.015803} (\bibinfo {year} {2010})\BibitemShut {NoStop}%
\bibitem [{\citenamefont {Grill}\ \emph {et~al.}(2014)\citenamefont {Grill}, \citenamefont {Pais}, \citenamefont {Provid\^encia}, \citenamefont {Vida\~na},\ and\ \citenamefont {Avancini}}]{Grill_2014}%
  \BibitemOpen
  \bibfield  {author} {\bibinfo {author} {\bibfnamefont {F.}~\bibnamefont {Grill}}, \bibinfo {author} {\bibfnamefont {H.}~\bibnamefont {Pais}}, \bibinfo {author} {\bibfnamefont {C.~m.~c.}\ \bibnamefont {Provid\^encia}}, \bibinfo {author} {\bibfnamefont {I.}~\bibnamefont {Vida\~na}},\ and\ \bibinfo {author} {\bibfnamefont {S.~S.}\ \bibnamefont {Avancini}},\ }\bibfield  {title} {\bibinfo {title} {{Equation of state and thickness of the inner crust of neutron stars}},\ }\href {https://doi.org/10.1103/PhysRevC.90.045803} {\bibfield  {journal} {\bibinfo  {journal} {Phys. Rev. C}\ }\textbf {\bibinfo {volume} {90}},\ \bibinfo {pages} {045803} (\bibinfo {year} {2014})}\BibitemShut {NoStop}%
\bibitem [{\citenamefont {Hannam}\ \emph {et~al.}(2014)\citenamefont {Hannam}, \citenamefont {Schmidt}, \citenamefont {Boh\'e}, \citenamefont {Haegel}, \citenamefont {Husa}, \citenamefont {Ohme}, \citenamefont {Pratten},\ and\ \citenamefont {P\"urrer}}]{Hannam_2014}%
  \BibitemOpen
  \bibfield  {author} {\bibinfo {author} {\bibfnamefont {M.}~\bibnamefont {Hannam}}, \bibinfo {author} {\bibfnamefont {P.}~\bibnamefont {Schmidt}}, \bibinfo {author} {\bibfnamefont {A.}~\bibnamefont {Boh\'e}}, \bibinfo {author} {\bibfnamefont {L.}~\bibnamefont {Haegel}}, \bibinfo {author} {\bibfnamefont {S.}~\bibnamefont {Husa}}, \bibinfo {author} {\bibfnamefont {F.}~\bibnamefont {Ohme}}, \bibinfo {author} {\bibfnamefont {G.}~\bibnamefont {Pratten}},\ and\ \bibinfo {author} {\bibfnamefont {M.}~\bibnamefont {P\"urrer}},\ }\bibfield  {title} {\bibinfo {title} {{Simple Model of Complete Precessing Black-Hole-Binary Gravitational Waveforms}},\ }\href {https://doi.org/10.1103/PhysRevLett.113.151101} {\bibfield  {journal} {\bibinfo  {journal} {Phys. Rev. Lett.}\ }\textbf {\bibinfo {volume} {113}},\ \bibinfo {pages} {151101} (\bibinfo {year} {2014})}\BibitemShut {NoStop}%
\bibitem [{GW1()}]{GW170817_data}%
  \BibitemOpen
  \href@noop {} {}\bibinfo {howpublished} {\url{https://dcc.ligo.org/LIGO-P1800061/public}}\BibitemShut {NoStop}%
\bibitem [{\citenamefont {Abac}\ \emph {et~al.}(2024{\natexlab{b}})\citenamefont {Abac}, \citenamefont {Dietrich}, \citenamefont {Buonanno}, \citenamefont {Steinhoff},\ and\ \citenamefont {Ujevic}}]{Abac_2024b}%
  \BibitemOpen
  \bibfield  {author} {\bibinfo {author} {\bibfnamefont {A.}~\bibnamefont {Abac}}, \bibinfo {author} {\bibfnamefont {T.}~\bibnamefont {Dietrich}}, \bibinfo {author} {\bibfnamefont {A.}~\bibnamefont {Buonanno}}, \bibinfo {author} {\bibfnamefont {J.}~\bibnamefont {Steinhoff}},\ and\ \bibinfo {author} {\bibfnamefont {M.}~\bibnamefont {Ujevic}},\ }\bibfield  {title} {\bibinfo {title} {{New and robust gravitational-waveform model for high-mass-ratio binary neutron star systems with dynamical tidal effects}},\ }\href {https://doi.org/10.1103/PhysRevD.109.024062} {\bibfield  {journal} {\bibinfo  {journal} {Phys. Rev. D}\ }\textbf {\bibinfo {volume} {109}},\ \bibinfo {pages} {024062} (\bibinfo {year} {2024}{\natexlab{b}})}\BibitemShut {NoStop}%
\bibitem [{\citenamefont {Matas}\ \emph {et~al.}(2020)\citenamefont {Matas}, \citenamefont {Dietrich}, \citenamefont {Buonanno}, \citenamefont {Hinderer}, \citenamefont {P\"urrer}, \citenamefont {Foucart}, \citenamefont {Boyle}, \citenamefont {Duez}, \citenamefont {Kidder}, \citenamefont {Pfeiffer},\ and\ \citenamefont {Scheel}}]{Matas_2020}%
  \BibitemOpen
  \bibfield  {author} {\bibinfo {author} {\bibfnamefont {A.}~\bibnamefont {Matas}}, \bibinfo {author} {\bibfnamefont {T.}~\bibnamefont {Dietrich}}, \bibinfo {author} {\bibfnamefont {A.}~\bibnamefont {Buonanno}}, \bibinfo {author} {\bibfnamefont {T.}~\bibnamefont {Hinderer}}, \bibinfo {author} {\bibfnamefont {M.}~\bibnamefont {P\"urrer}}, \bibinfo {author} {\bibfnamefont {F.}~\bibnamefont {Foucart}}, \bibinfo {author} {\bibfnamefont {M.}~\bibnamefont {Boyle}}, \bibinfo {author} {\bibfnamefont {M.~D.}\ \bibnamefont {Duez}}, \bibinfo {author} {\bibfnamefont {L.~E.}\ \bibnamefont {Kidder}}, \bibinfo {author} {\bibfnamefont {H.~P.}\ \bibnamefont {Pfeiffer}},\ and\ \bibinfo {author} {\bibfnamefont {M.~A.}\ \bibnamefont {Scheel}},\ }\bibfield  {title} {\bibinfo {title} {Aligned-spin neutron-star--black-hole waveform model based on the effective-one-body approach and numerical-relativity simulations},\ }\href {https://doi.org/10.1103/PhysRevD.102.043023} {\bibfield  {journal} {\bibinfo  {journal} {Phys. Rev. D}\
  }\textbf {\bibinfo {volume} {102}},\ \bibinfo {pages} {043023} (\bibinfo {year} {2020})}\BibitemShut {NoStop}%
\bibitem [{\citenamefont {Blanchet}(2024)}]{Blanchet_2024}%
  \BibitemOpen
  \bibfield  {author} {\bibinfo {author} {\bibfnamefont {L.}~\bibnamefont {Blanchet}},\ }\bibfield  {title} {\bibinfo {title} {Constraining neutron-star tidal love numbers with gravitational-wave detectors},\ }\bibfield  {journal} {\bibinfo  {journal} {Living Rev Relativ}\ }\href {https://doi.org/10.1007/s41114-024-00050-z} {10.1007/s41114-024-00050-z} (\bibinfo {year} {2024})\BibitemShut {NoStop}%
\bibitem [{\citenamefont {Babak}\ \emph {et~al.}(2013)\citenamefont {Babak}, \citenamefont {Biswas}, \citenamefont {Brady}, \citenamefont {Brown}, \citenamefont {Cannon}, \citenamefont {Capano}, \citenamefont {Clayton}, \citenamefont {Cokelaer}, \citenamefont {Creighton}, \citenamefont {Dent}, \citenamefont {Dietz}, \citenamefont {Fairhurst}, \citenamefont {Fotopoulos}, \citenamefont {Gonz\'alez}, \citenamefont {Hanna}, \citenamefont {Harry}, \citenamefont {Jones}, \citenamefont {Keppel}, \citenamefont {McKechan}, \citenamefont {Pekowsky}, \citenamefont {Privitera}, \citenamefont {Robinson}, \citenamefont {Rodriguez}, \citenamefont {Sathyaprakash}, \citenamefont {Sengupta}, \citenamefont {Vallisneri}, \citenamefont {Vaulin},\ and\ \citenamefont {Weinstein}}]{Babak_2006}%
  \BibitemOpen
  \bibfield  {author} {\bibinfo {author} {\bibfnamefont {S.}~\bibnamefont {Babak}}, \bibinfo {author} {\bibfnamefont {R.}~\bibnamefont {Biswas}}, \bibinfo {author} {\bibfnamefont {P.~R.}\ \bibnamefont {Brady}}, \bibinfo {author} {\bibfnamefont {D.~A.}\ \bibnamefont {Brown}}, \bibinfo {author} {\bibfnamefont {K.}~\bibnamefont {Cannon}}, \bibinfo {author} {\bibfnamefont {C.~D.}\ \bibnamefont {Capano}}, \bibinfo {author} {\bibfnamefont {J.~H.}\ \bibnamefont {Clayton}}, \bibinfo {author} {\bibfnamefont {T.}~\bibnamefont {Cokelaer}}, \bibinfo {author} {\bibfnamefont {J.~D.~E.}\ \bibnamefont {Creighton}}, \bibinfo {author} {\bibfnamefont {T.}~\bibnamefont {Dent}}, \bibinfo {author} {\bibfnamefont {A.}~\bibnamefont {Dietz}}, \bibinfo {author} {\bibfnamefont {S.}~\bibnamefont {Fairhurst}}, \bibinfo {author} {\bibfnamefont {N.}~\bibnamefont {Fotopoulos}}, \bibinfo {author} {\bibfnamefont {G.}~\bibnamefont {Gonz\'alez}}, \bibinfo {author} {\bibfnamefont {C.}~\bibnamefont {Hanna}}, \bibinfo {author} {\bibfnamefont
  {I.~W.}\ \bibnamefont {Harry}}, \bibinfo {author} {\bibfnamefont {G.}~\bibnamefont {Jones}}, \bibinfo {author} {\bibfnamefont {D.}~\bibnamefont {Keppel}}, \bibinfo {author} {\bibfnamefont {D.~J.~A.}\ \bibnamefont {McKechan}}, \bibinfo {author} {\bibfnamefont {L.}~\bibnamefont {Pekowsky}}, \bibinfo {author} {\bibfnamefont {S.}~\bibnamefont {Privitera}}, \bibinfo {author} {\bibfnamefont {C.}~\bibnamefont {Robinson}}, \bibinfo {author} {\bibfnamefont {A.~C.}\ \bibnamefont {Rodriguez}}, \bibinfo {author} {\bibfnamefont {B.~S.}\ \bibnamefont {Sathyaprakash}}, \bibinfo {author} {\bibfnamefont {A.~S.}\ \bibnamefont {Sengupta}}, \bibinfo {author} {\bibfnamefont {M.}~\bibnamefont {Vallisneri}}, \bibinfo {author} {\bibfnamefont {R.}~\bibnamefont {Vaulin}},\ and\ \bibinfo {author} {\bibfnamefont {A.~J.}\ \bibnamefont {Weinstein}},\ }\bibfield  {title} {\bibinfo {title} {Searching for gravitational waves from binary coalescence},\ }\href {https://doi.org/10.1103/PhysRevD.87.024033} {\bibfield  {journal} {\bibinfo
  {journal} {Phys. Rev. D}\ }\textbf {\bibinfo {volume} {87}},\ \bibinfo {pages} {024033} (\bibinfo {year} {2013})}\BibitemShut {NoStop}%
\bibitem [{\citenamefont {Brown}\ \emph {et~al.}(2012)\citenamefont {Brown}, \citenamefont {Harry}, \citenamefont {Lundgren},\ and\ \citenamefont {Nitz}}]{Brown_2012}%
  \BibitemOpen
  \bibfield  {author} {\bibinfo {author} {\bibfnamefont {D.~A.}\ \bibnamefont {Brown}}, \bibinfo {author} {\bibfnamefont {I.}~\bibnamefont {Harry}}, \bibinfo {author} {\bibfnamefont {A.}~\bibnamefont {Lundgren}},\ and\ \bibinfo {author} {\bibfnamefont {A.~H.}\ \bibnamefont {Nitz}},\ }\bibfield  {title} {\bibinfo {title} {{Detecting binary neutron star systems with spin in advanced gravitational-wave detectors}},\ }\href {https://doi.org/10.1103/PhysRevD.86.084017} {\bibfield  {journal} {\bibinfo  {journal} {Phys. Rev. D}\ }\textbf {\bibinfo {volume} {86}},\ \bibinfo {pages} {084017} (\bibinfo {year} {2012})}\BibitemShut {NoStop}%
\bibitem [{\citenamefont {Harry}\ \emph {et~al.}(2014)\citenamefont {Harry}, \citenamefont {Nitz}, \citenamefont {Brown}, \citenamefont {Lundgren}, \citenamefont {Ochsner},\ and\ \citenamefont {Keppel}}]{Harry_2014}%
  \BibitemOpen
  \bibfield  {author} {\bibinfo {author} {\bibfnamefont {I.~W.}\ \bibnamefont {Harry}}, \bibinfo {author} {\bibfnamefont {A.~H.}\ \bibnamefont {Nitz}}, \bibinfo {author} {\bibfnamefont {D.~A.}\ \bibnamefont {Brown}}, \bibinfo {author} {\bibfnamefont {A.~P.}\ \bibnamefont {Lundgren}}, \bibinfo {author} {\bibfnamefont {E.}~\bibnamefont {Ochsner}},\ and\ \bibinfo {author} {\bibfnamefont {D.}~\bibnamefont {Keppel}},\ }\bibfield  {title} {\bibinfo {title} {Investigating the effect of precession on searches for neutron-star--black-hole binaries with advanced ligo},\ }\href {https://doi.org/10.1103/PhysRevD.89.024010} {\bibfield  {journal} {\bibinfo  {journal} {Phys. Rev. D}\ }\textbf {\bibinfo {volume} {89}},\ \bibinfo {pages} {024010} (\bibinfo {year} {2014})}\BibitemShut {NoStop}%
\bibitem [{\citenamefont {Hanna}\ \emph {et~al.}(2023)\citenamefont {Hanna} \emph {et~al.}}]{Hanna_2022}%
  \BibitemOpen
  \bibfield  {author} {\bibinfo {author} {\bibfnamefont {C.}~\bibnamefont {Hanna}} \emph {et~al.},\ }\bibfield  {title} {\bibinfo {title} {{Binary tree approach to template placement for searches for gravitational waves from compact binary mergers}},\ }\href {https://doi.org/10.1103/PhysRevD.108.042003} {\bibfield  {journal} {\bibinfo  {journal} {Phys. Rev. D}\ }\textbf {\bibinfo {volume} {108}},\ \bibinfo {pages} {042003} (\bibinfo {year} {2023})},\ \Eprint {https://arxiv.org/abs/2209.11298} {arXiv:2209.11298 [gr-qc]} \BibitemShut {NoStop}%
\bibitem [{\citenamefont {Phukon}\ \emph {et~al.}(2025)\citenamefont {Phukon}, \citenamefont {Schmidt},\ and\ \citenamefont {Pratten}}]{Phukon:2024amh}%
  \BibitemOpen
  \bibfield  {author} {\bibinfo {author} {\bibfnamefont {K.~S.}\ \bibnamefont {Phukon}}, \bibinfo {author} {\bibfnamefont {P.}~\bibnamefont {Schmidt}},\ and\ \bibinfo {author} {\bibfnamefont {G.}~\bibnamefont {Pratten}},\ }\bibfield  {title} {\bibinfo {title} {{Geometric template bank for the detection of spinning low-mass compact binaries with moderate orbital eccentricity}},\ }\href {https://doi.org/10.1103/PhysRevD.111.043040} {\bibfield  {journal} {\bibinfo  {journal} {Phys. Rev. D}\ }\textbf {\bibinfo {volume} {111}},\ \bibinfo {pages} {043040} (\bibinfo {year} {2025})},\ \Eprint {https://arxiv.org/abs/2412.06433} {arXiv:2412.06433 [gr-qc]} \BibitemShut {NoStop}%
\bibitem [{\citenamefont {Babak}(2008)}]{Babak_2008}%
  \BibitemOpen
  \bibfield  {author} {\bibinfo {author} {\bibfnamefont {S.}~\bibnamefont {Babak}},\ }\bibfield  {title} {\bibinfo {title} {Building a stochastic template bank for detecting massive black hole binaries},\ }\href {https://doi.org/10.1088/0264-9381/25/19/195011} {\bibfield  {journal} {\bibinfo  {journal} {Classical and Quantum Gravity}\ }\textbf {\bibinfo {volume} {25}},\ \bibinfo {pages} {195011} (\bibinfo {year} {2008})}\BibitemShut {NoStop}%
\bibitem [{\citenamefont {Ajith}\ \emph {et~al.}(2014)\citenamefont {Ajith}, \citenamefont {Fotopoulos}, \citenamefont {Privitera}, \citenamefont {Neunzert}, \citenamefont {Mazumder},\ and\ \citenamefont {Weinstein}}]{Ajith_2014}%
  \BibitemOpen
  \bibfield  {author} {\bibinfo {author} {\bibfnamefont {P.}~\bibnamefont {Ajith}}, \bibinfo {author} {\bibfnamefont {N.}~\bibnamefont {Fotopoulos}}, \bibinfo {author} {\bibfnamefont {S.}~\bibnamefont {Privitera}}, \bibinfo {author} {\bibfnamefont {A.}~\bibnamefont {Neunzert}}, \bibinfo {author} {\bibfnamefont {N.}~\bibnamefont {Mazumder}},\ and\ \bibinfo {author} {\bibfnamefont {A.~J.}\ \bibnamefont {Weinstein}},\ }\bibfield  {title} {\bibinfo {title} {Effectual template bank for the detection of gravitational waves from inspiralling compact binaries with generic spins},\ }\href {https://doi.org/10.1103/PhysRevD.89.084041} {\bibfield  {journal} {\bibinfo  {journal} {Phys. Rev. D}\ }\textbf {\bibinfo {volume} {89}},\ \bibinfo {pages} {084041} (\bibinfo {year} {2014})}\BibitemShut {NoStop}%
\bibitem [{\citenamefont {Capano}\ \emph {et~al.}(2016)\citenamefont {Capano}, \citenamefont {Harry}, \citenamefont {Privitera},\ and\ \citenamefont {Buonanno}}]{Capanno_2016}%
  \BibitemOpen
  \bibfield  {author} {\bibinfo {author} {\bibfnamefont {C.}~\bibnamefont {Capano}}, \bibinfo {author} {\bibfnamefont {I.}~\bibnamefont {Harry}}, \bibinfo {author} {\bibfnamefont {S.}~\bibnamefont {Privitera}},\ and\ \bibinfo {author} {\bibfnamefont {A.}~\bibnamefont {Buonanno}},\ }\bibfield  {title} {\bibinfo {title} {Implementing a search for gravitational waves from binary black holes with nonprecessing spin},\ }\href {https://doi.org/10.1103/PhysRevD.93.124007} {\bibfield  {journal} {\bibinfo  {journal} {Phys. Rev. D}\ }\textbf {\bibinfo {volume} {93}},\ \bibinfo {pages} {124007} (\bibinfo {year} {2016})}\BibitemShut {NoStop}%
\bibitem [{\citenamefont {Fehrmann}\ and\ \citenamefont {Pletsch}(2014)}]{Fehrmann_2014}%
  \BibitemOpen
  \bibfield  {author} {\bibinfo {author} {\bibfnamefont {H.}~\bibnamefont {Fehrmann}}\ and\ \bibinfo {author} {\bibfnamefont {H.~J.}\ \bibnamefont {Pletsch}},\ }\bibfield  {title} {\bibinfo {title} {Efficient generation and optimization of stochastic template banks by a neighboring cell algorithm},\ }\href {https://doi.org/10.1103/PhysRevD.90.124049} {\bibfield  {journal} {\bibinfo  {journal} {Phys. Rev. D}\ }\textbf {\bibinfo {volume} {90}},\ \bibinfo {pages} {124049} (\bibinfo {year} {2014})}\BibitemShut {NoStop}%
\bibitem [{\citenamefont {Owen}\ and\ \citenamefont {Sathyaprakash}(1999)}]{Owen_1999}%
  \BibitemOpen
  \bibfield  {author} {\bibinfo {author} {\bibfnamefont {B.~J.}\ \bibnamefont {Owen}}\ and\ \bibinfo {author} {\bibfnamefont {B.~S.}\ \bibnamefont {Sathyaprakash}},\ }\bibfield  {title} {\bibinfo {title} {{Matched filtering of gravitational waves from inspiraling compact binaries: Computational cost and template placement}},\ }\bibfield  {journal} {\bibinfo  {journal} {Physical Review D}\ }\textbf {\bibinfo {volume} {60}},\ \href {https://doi.org/10.1103/physrevd.60.022002} {10.1103/physrevd.60.022002} (\bibinfo {year} {1999})\BibitemShut {NoStop}%
\bibitem [{\citenamefont {Cokelaer}(2007)}]{Cokelaer_2007}%
  \BibitemOpen
  \bibfield  {author} {\bibinfo {author} {\bibfnamefont {T.}~\bibnamefont {Cokelaer}},\ }\bibfield  {title} {\bibinfo {title} {{Gravitational waves from inspiralling compact binaries: Hexagonal template placement and its efficiency in detecting physical signals}},\ }\bibfield  {journal} {\bibinfo  {journal} {Physical Review D}\ }\textbf {\bibinfo {volume} {76}},\ \href {https://doi.org/10.1103/physrevd.76.102004} {10.1103/physrevd.76.102004} (\bibinfo {year} {2007})\BibitemShut {NoStop}%
\bibitem [{\citenamefont {Sharma}\ \emph {et~al.}(2024)\citenamefont {Sharma}, \citenamefont {Roy},\ and\ \citenamefont {Sengupta}}]{Sharma:2023djw}%
  \BibitemOpen
  \bibfield  {author} {\bibinfo {author} {\bibfnamefont {A.}~\bibnamefont {Sharma}}, \bibinfo {author} {\bibfnamefont {S.}~\bibnamefont {Roy}},\ and\ \bibinfo {author} {\bibfnamefont {A.~S.}\ \bibnamefont {Sengupta}},\ }\bibfield  {title} {\bibinfo {title} {{Template bank to search for exotic gravitational wave signals from astrophysical compact binaries}},\ }\href {https://doi.org/10.1103/PhysRevD.109.124049} {\bibfield  {journal} {\bibinfo  {journal} {Phys. Rev. D}\ }\textbf {\bibinfo {volume} {109}},\ \bibinfo {pages} {124049} (\bibinfo {year} {2024})},\ \Eprint {https://arxiv.org/abs/2311.03274} {arXiv:2311.03274 [gr-qc]} \BibitemShut {NoStop}%
\bibitem [{\citenamefont {Parzen}(1962)}]{Parzen_1962}%
  \BibitemOpen
  \bibfield  {author} {\bibinfo {author} {\bibfnamefont {E.}~\bibnamefont {Parzen}},\ }\bibfield  {title} {\bibinfo {title} {On estimation of a probability density function and mode},\ }\href {http://www.jstor.org/stable/2237880} {\bibfield  {journal} {\bibinfo  {journal} {The Annals of Mathematical Statistics}\ }\textbf {\bibinfo {volume} {33}},\ \bibinfo {pages} {1065} (\bibinfo {year} {1962})}\BibitemShut {NoStop}%
\bibitem [{\citenamefont {Falxa}\ \emph {et~al.}(2023)\citenamefont {Falxa}, \citenamefont {Babak},\ and\ \citenamefont {Le~Jeune}}]{Falxa_2023}%
  \BibitemOpen
  \bibfield  {author} {\bibinfo {author} {\bibfnamefont {M.}~\bibnamefont {Falxa}}, \bibinfo {author} {\bibfnamefont {S.}~\bibnamefont {Babak}},\ and\ \bibinfo {author} {\bibfnamefont {M.}~\bibnamefont {Le~Jeune}},\ }\bibfield  {title} {\bibinfo {title} {Adaptive kernel density estimation proposal in gravitational wave data analysis},\ }\href {https://doi.org/10.1103/PhysRevD.107.022008} {\bibfield  {journal} {\bibinfo  {journal} {Phys. Rev. D}\ }\textbf {\bibinfo {volume} {107}},\ \bibinfo {pages} {022008} (\bibinfo {year} {2023})}\BibitemShut {NoStop}%
\end{thebibliography}%

\end{document}